\documentclass[journal]{new-aiaa}
\usepackage[utf8]{inputenc}
\usepackage{textcomp}
\usepackage{graphicx}
\usepackage{amsmath}
\usepackage[version=4]{mhchem}
\usepackage{siunitx}
\usepackage{longtable,tabularx}

\title{An Automatic Tree Search Algorithm for the Tisserand Graph}

\author{David de la Torre Sangr\`a}
\affil{Polytechnic University of Catalonia, Physics Department, ESEIAAT, Colom 11, 08222 Terrassa, Spain.}
\author{Elena Fantino\footnote{Corresponding author: elena.fantino@ku.ac.ae}}
\affil{Department of Aerospace Engineering, Khalifa University of Science and Technology, P.O. Box 127788, Abu Dhabi, United Arab Emirates.}
\author{Roberto Flores}
\affil{Department of Aerospace Engineering, Khalifa University of Science and Technology, P.O. Box 127788, Abu Dhabi, United Arab Emirates.\\
Centre Internacional de M\`etodes Num\`erics en Enginyeria CIMNE, 08034 Barcelona, Spain.}
\author{Oscar Calvente Lozano and Celestino Garc\'ia Estelrich} 
\affil{Polytechnic University of Catalonia, Physics Department, ESEIAAT, Colom 11, 08222 Terrassa, Spain.}

\begin{document}

\maketitle

\begin{abstract}
The Tisserand graph (TG) is a graphical tool commonly employed in the preliminary design of gravity-assisted trajectories. The TG is a two-dimensional map showing essential orbital information regarding the Keplerian orbits resulting from the close passage by one or more  massive bodies, given the magnitude of the hyperbolic excess speed ($v_{\infty}$) and the minimum allowed pericenter height for each passage. Contours of constant $v_{\infty}$ populate the TG. Intersections between contours allow to link consecutive flybys and build sequences of encounters en route to a selected destination.
When the number of perturbing bodies is large and many $v_{\infty}$ levels are considered, the identification of all the possible sequences of encounters through the visual inspection of the TG  becomes a laborious task.  Besides, if the sequences are used as input for a numerical code for trajectory design and optimization, an automated examination of the TG is desirable. 
This contribution describes an automatic technique to explore the TG and find all the encounter paths. The technique is based on a tree search method, and the intersections between contours are found using the regula-falsi scheme. 
The method is validated through comparisons with solutions available in the open literature. Examples are given of application to interplanetary mission scenarios, including the coupling with a trajectory optimizer. 
\end{abstract}

\vspace{0.5cm}

\section{Introduction}
\lettrine{T}he Tisserand graph (TG) is a graphical tool used in the preliminary design of gravity-assisted trajectories. 
By displaying essential orbital information about the Keplerian orbits resulting from  close passages to a set of massive bodies, the TG helps construct a sequence of encounters between a starting and a destination orbit. 

The TG is named after $19^{th}$ century astronomer Fran\c{c}ois F\'elix Tisserand, who developed a method -the Tisserand's criterion \cite{Roy2005}- to identify an object (a comet or an asteroid) after a passage by a planet. The orbital elements of the object may change after the close approach, but the Tisserand parameter, a function of semi-major axis, eccentricity and inclination of the orbit, stays approximately constant and can be used to identify the object after the event. The outcome of the flyby with a planet is shown in the TG as curves, the $v_{\infty}$ contours, corresponding to all hyperbolic passages with a given excess speed $v_{\infty}$. This velocity is closely related to the Tisserand parameter. Each point along the contour corresponds to the angle between the hyperbolic excess velocity of the object and the velocity of the body (see below).
The intersections between contours link encounters with different planets, hence they can be used to build paths to a selected destination. 

\begin{figure}[htb]
\centering
\includegraphics[width=4.0in]{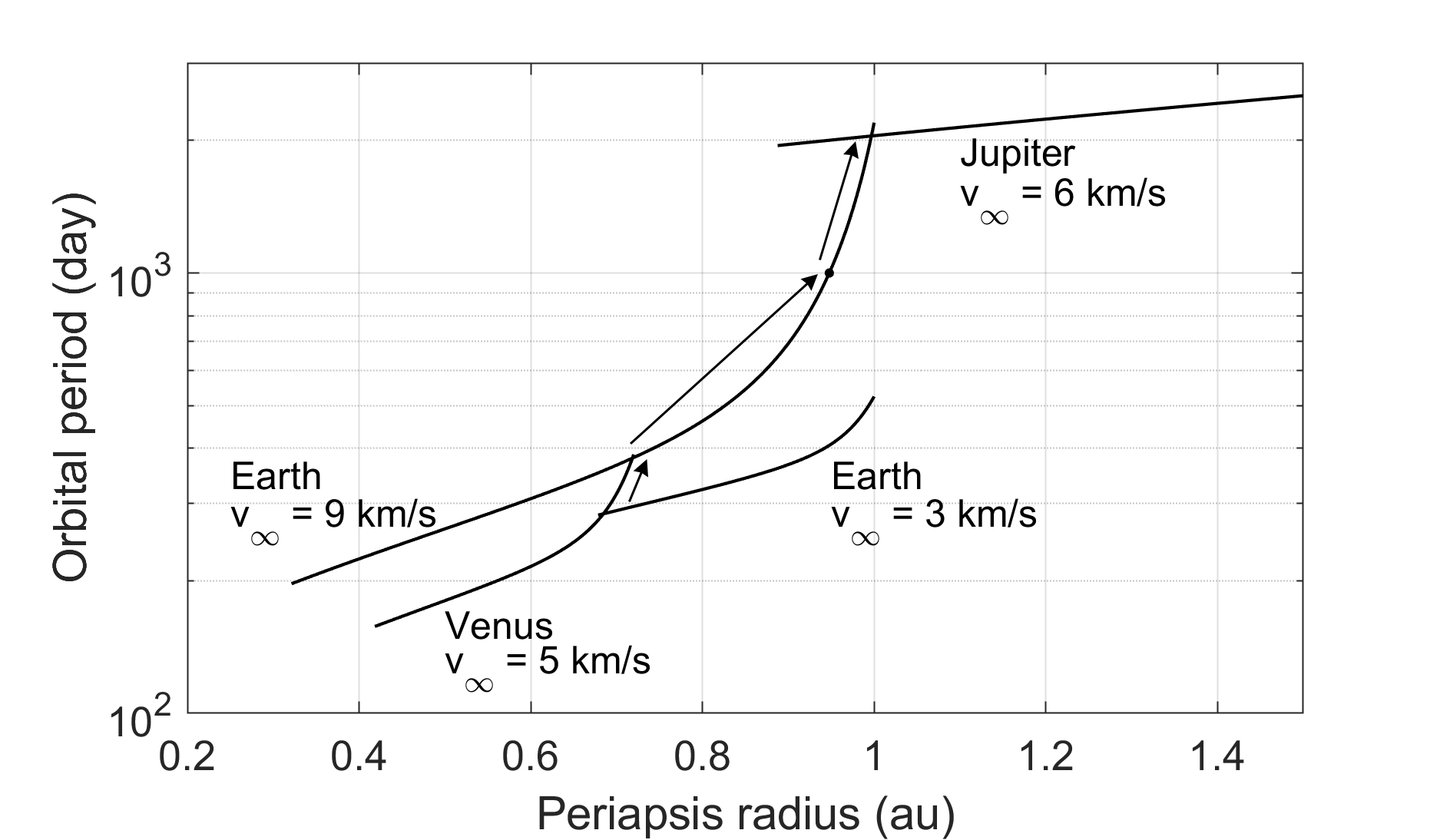}
\caption{TG for an Earth-to-Jupiter trajectory with intermediate Earth and Venus flybys \cite{Strange2002}.}
\label{fig:Strange1}
\end{figure}

Figure~\ref{fig:Strange1} illustrates a TG for an Earth-to-Jupiter trajectory in which Earth and Venus flybys are the options considered ~\cite{Strange2002}. The map shows the orbital periods and the perihelia of the heliocentric orbits that a spacecraft (S/C) can follow as a result of gravity assists with Earth (with $v_{\infty}$ of 3 and 9 km/s) and Venus ($v_{\infty}$ of 6 km/s) before approaching Jupiter at a relative speed of 6 km/s.  The intersections between contours yield the following sequence of encounters: Earth (launch), Venus, Earth, Earth, Jupiter (arrival). 
The basic assumption of the TG is that all the planets are on circular coplanar orbits. The planet positions are not  considered, i.e., flybys are assumed to occur whenever the orbit of the S/C intersects that of a planet. For this reason, this tool is confined to a preliminary mission design stage, which must be followed by an analysis considering the phasing constraints.

The TG has been employed in interplanetary trajectory design for many years. \cite{Strange2002} discussed  Tisserand's theory in great depth, applied the TG to a wide number of transfers and highlighted the importance of an automatic exploration of the graph for complex scenarios. \cite{Miller2002} reviewed the theoretical background of Tisserand's criterion and illustrated its application to the preliminary design of Cassini's interplanetary trajectory. \cite{Heaton2002} used the TG in the design of tours in the Jovian system for the Europa Orbiter mission. They found sequences of lunar encounters (Europa, Ganymede, Callisto) which were then input to the Satellite Tour Design Program \cite{Diehl1983} designed for the Galileo S/C by JPL. That work emphasized the importance of an automatic method to search for transfers within the TG. \cite{Heaton2003} designed a tour of the Uranus system using the TG to adjust the inclination of the target science orbit around Ariel. \cite{Okutsu2002} employed a TG to design Mars free-return trajectories via gravity assists with Venus. \cite{Khan2004} conducted the mission analysis for a two-S/C (relay and orbiter) low-cost mission to Europa, in which the TG is used to identify tour options for both vehicles: a tour in the inner Jovian system for the orbiter and a tour of the outer, radiation-safe system for the relay. \cite{Campagnola2010a,Strange2009} derived a new formulation for $v_\infty$ leveraging maneuvers (VILMs) within the so-called Tisserand Leveraging Graph, used as a tool to design endgames. The new method allows rapid calculations of the minimum useful $\Delta V$ using VILMs to design resonant lunar tours at Jupiter and Saturn. \cite{Campagnola2010C} worked out a linear approximation to the solution space allowing fast sequence searches, and used the methodology to design a trajectory for an Enceladus orbiter. Then,  \cite{Campagnola2010B, Campagnola2012} extended the formulation of the TG to the circular restricted three-body problem (CR3BP). They found a trajectory encountering Callisto that inserts the S/C into a circular orbit around Europa, improving by 30\% the $\Delta V$ budget of the classical patched-conics method.
\cite{Kloster2011} presented a design of a Jovian tour for an Europa orbiter mission in which 
the TG is used in combination with a simple radiation model to avoid hazardous exposures during the flybys.
\cite{Lantoine2011} relied on the Tisserand-Poincar\'e (T-P) graph, a variant of the TG for the CR3BP, to obtain initial guesses of inter-moon transfers in the Jovian system in a patched three-body model.
\cite{Hughes2013} investigated a broad collection of ballistic trajectories to Neptune using the TG for the selection of the planetary encounters. The trajectory was solved by patched conics with impulsive manoeuvres either in the form of powered gravity-assists or with VILMs, i.e., using deep-space manoeuvres to lower the launch $\Delta V$. The authors highlighted the effects of phasing and mission constraints and estimated that only 21 out of 76 encounter sequences were feasible.
\cite{Strange2014} applied a TG-based method to the CR3BP to identify ways of capturing small asteroids around the Earth redirecting them to lunar gravity assists by means of a small ($<$ 200 m/s) $\Delta V$. 
\cite{Colasurdo2014} employed the TG to design a tour of the Galilean moons using resonant transfers, which was the winning solution of the $6^{th}$ edition of the Global Trajectory Optimization Competition.
\cite{Campagnola2014a} investigated three Jovian tour mission configurations (flyby-only, orbiter and lander) using the T-P graph. 
The solution achieves low $\Delta V$ by means of high-altitude flybys and deep-space manoeuvres.
\cite{Maiwald2016} adapted the TG to a low-thrust mission: the variation in the orbital energy over a thrust arc corresponds to a jump between different $v_{\infty}$ contours in the TG.
\cite{Yarnoz2016} developed a systematic approach to generate multiple lunar flyby sequences for small interplanetary probes in a CR3BP, using the third-body perturbation of the Sun as a VILM equivalent.
\cite{Jones2017} employed the TG to study the triple cycler family of orbits among Earth, Mars and Venus. This type of trajectories periodically cycle between flybys of Venus, Earth, and Mars and were conceived for future manned mission to Mars. The solutions are characterized by lower $\Delta V$ requirements than traditional Earth-Mars cyclers.

\begin{figure}[htb]
\centering
\includegraphics[width=5.0in]{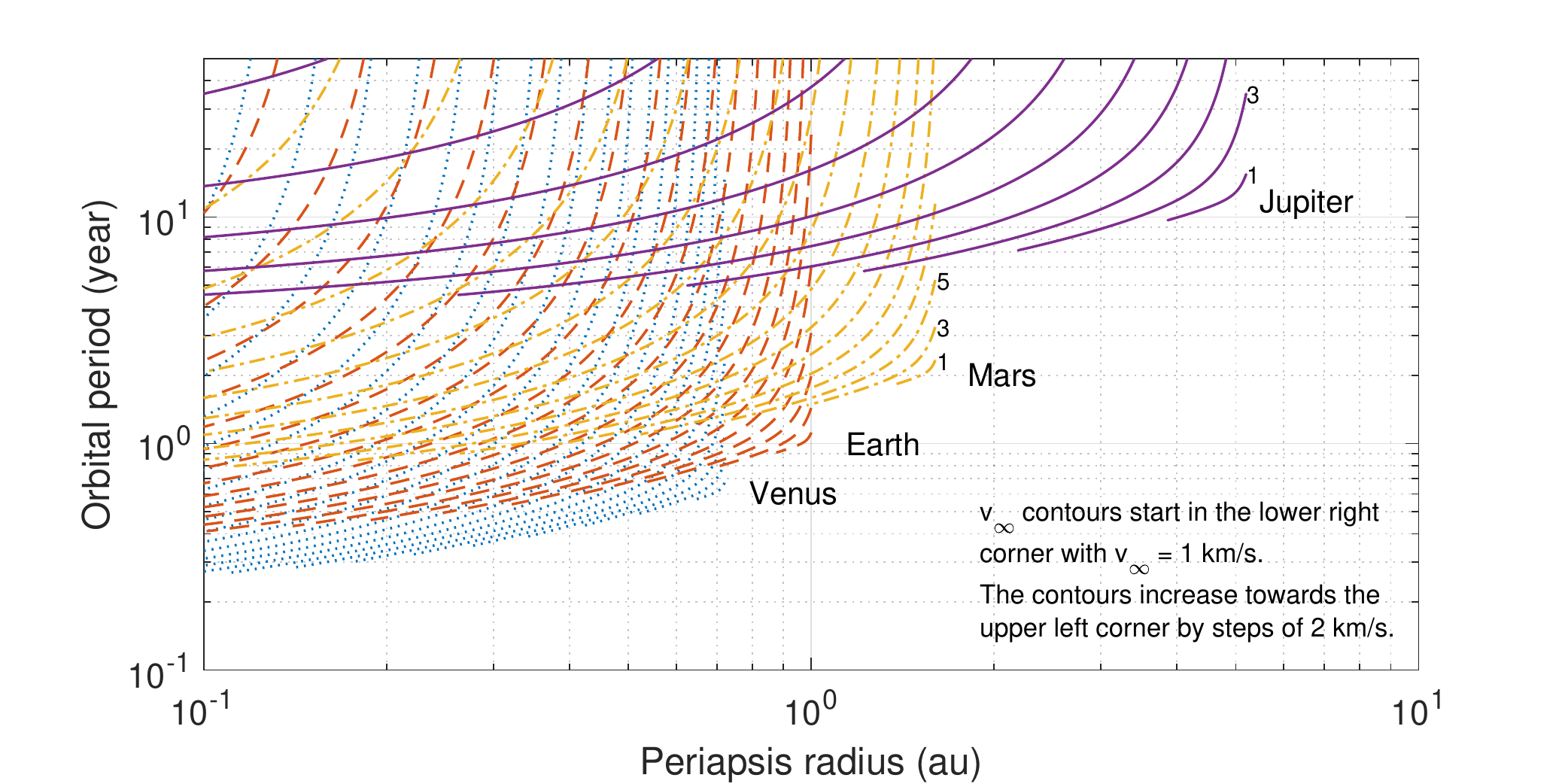}
\caption{TG for an Earth-to-Jupiter trajectory in which flybys with Earth, Venus and Mars are considered \cite{Strange2002}.}
\label{fig:Strange2}
\end{figure}

When the number of planets and $v_{\infty}$ levels increases (see, e.g., Fig.~\ref{fig:Strange2}), the identification of all the possible sequences of encounters by visual inspection becomes impractical (see also \cite{Strange2002}). Besides, if the sequence of encounters is used as input of a numerical code for trajectory design and optimization, an automated examination of the TG is desirable. 
Automated strategies for analysis of the TG are in widespread use, but, to the best of the authors' knowledge, they have not been documented in public literature. This contribution describes the development, implementation, validation and application of an automatic technique to explore the TG and determine all the sequences of flybys that it contains. The technique is based on a tree search method and will be referred to as the Tisserand PathFinder (TPF) algorithm. 

Section~\ref{sec:theory} reviews the derivation of the Tisserand parameter, while Sect.~\ref{sec:TG} illustrates the procedure for the construction of a TG and gives simple guidelines for selecting ranges of $v_{\infty}$ levels to include in a TG.
Section~\ref{sec:tpf} illustrates the approach adopted to find intersections between contours, build a tree, explore it and construct sequences of encounters. Section~\ref{sec:test} presents a set of validation tests and Sect.~\ref{sec:app} provides two application examples. The conclusions are laid out in Sect.~\ref{sec:conclusions}. 

\section{Tisserand's parameter}
\label{sec:theory}
Tisserand's parameter is defined in a system of three bodies, two of which (the primaries) are assumed in circular orbits about each other, and the third body has negligible mass. This is the framework of the CR3BP. This dynamical model admits a constant of motion for the third body, i.e., Jacobi's integral $C_J$ \cite{Roy2005}, 
\begin{equation}
C_J = n^2\left(x^2+y^2\right) + 2\left(\frac{\mu_1}{r_1} + \frac{\mu_2}{r_2}\right) - \left(\dot{x}^2 + \dot{y}^2 + \dot{z}^2\right),
\label{eq:Tiss}
\end{equation}
in which $n$ is the mean motion of the primaries, $\mu_1$ and $\mu_2$ are their standard gravitational parameters and $r_1$ and $r_2$ are the distances from the third body to the two primaries. The reference frame is synodic (rotating, with the primaries fixed on the $x$-axis, the $xy$-plane being their orbital plane) and barycentric (with origin at the center of mass of the primaries). With respect to a barycentric inertial reference frame with axes $\xi$, $\eta$, $\zeta$ initially parallel to the synodic axes, Eq.~\ref{eq:Tiss} transforms into
\begin{equation}
C_J = 2n\left(\xi \dot{\eta} - \eta \dot{\xi}\right) + 2 \left({\frac{\mu_1}{r_1}}+{\frac{\mu_2}{r_2}}\right) - \left({\dot{\xi}}^2 + {\dot{\eta}}^2 + {\dot{\zeta}}^2\right).
\label{eq:tiss1_2}
\end{equation}
Assuming that the first primary is the Sun and the second one is a planet, $(\mu_2/\mu_1)^{2/5} \ll 1$. Hence, the center of mass of the system is very close to the Sun, and the radius of the sphere of influence  of the planet (see, e.g., \cite{Kaplan1976} Chapter 7, pp. 287-289) is much smaller than the distance $a_{2}$ between primaries. Therefore, when the third body (the S/C) approaches the planet, it is possible to approximate $r_1 \simeq a_{2}$ and $\mu_2/r_2 \simeq 0$ before entering the sphere of influence.
Under these conditions, the heliocentric trajectory of the S/C is approximately Keplerian, and $C_J$ can be rewritten in terms of the orbital elements. In particular, the last term of Eq.~\ref{eq:tiss1_2} can be approximated by the square of the Keplerian velocity $v$ of the S/C
\begin{equation}
{\dot{\xi}}^2 + {\dot{\eta}}^2 + {\dot{\zeta}}^2 \simeq   v^2 = \mu_1 \left(\frac{2}{r_1} - \frac{1}{a}\right) 
\simeq \mu_1 \left(\frac{2}{a_2} - \frac{1}{a}\right),
\label{eq:tiss2}
\end{equation}
with $a$ the S/C's semi-major axis. 
Then, the first term on the right-hand side of Eq.~\ref{eq:tiss1_2} can be rewritten by recalling that 
\begin{equation}
\xi \dot{\eta} - \eta \dot{\xi} = h \cos{i},
\label{eq:tiss3}
\end{equation}
with $h$ the magnitude of the specific orbital angular momentum and $i$ the orbital inclination.
Furthermore, 
\begin{equation}
h = \sqrt{\mu_1 a(1 - e^2)},
\label{eq:tiss4}
\end{equation}
$e$ being the orbital eccentricity. Hence
\begin{equation}
C_J \simeq  2n\sqrt{\mu_1 a(1 - e^2)} \cos i + \frac{\mu_1}{a}.
\label{eq:tiss1_3}
\end{equation}
In dimensionless variables, using  $\mu_1$ and $a_2$ as reference magnitudes, the mean motion and heliocentric velocity $v_2$ of the planet have unit value. The expression of the normalized Jacobi's integral $\bar{C}_J$ (hereinafter, dimensionless quantities will be denoted by barred symbols) reads
\begin{equation}
\bar{C}_J  \simeq 2\sqrt{\bar{a}(1-e^2)} \cos{i} + \frac{1}{\bar{a}}.
\label{eq:tiss5}
\end{equation}
The right-hand side of the above equation is Tisserand's parameter $\bar{C}_T$ \cite{Murray2000}.

As the S/C approaches the planet, the magnitude $\bar{v}_{\infty}$ of the relative velocity  $\bar{\bf v}_{\infty}$ can be obtained by applying the law of cosines to the triangle formed by the heliocentric velocity of the S/C
($\bar{\bf v}$) and planet ($\bar{\bf v}_2$):
\begin{equation}
\bar{v}_{\infty}^2 = \bar{v}^2 + 1 - 2 \bar{v} \cos{\gamma} \cos{i},
\label{eq:tiss6}
\end{equation}
where $\gamma$ is the S/C flight path angle and $\cos{\gamma} \cos{i}$ is the cosine of the angle between $\bar{\bf v}$ and $\bar{\bf v}_2$. Furthermore, 
\begin{equation}
\bar{v} \cos{\gamma} = \bar{h}.
\label{eq:h1}
\end{equation}
which, by means of Eq.~\ref{eq:tiss4}, provides
\begin{equation}
\bar{v}\cos{\gamma} = \sqrt{\bar{a}(1 - e^2)}.
\label{eq:tiss7}
\end{equation}
Recalling Eq.~\ref{eq:tiss2} and substituting Eq.~\ref{eq:tiss7} into Eq.~\ref{eq:tiss6} yields
\begin{equation}
\bar{v}_\infty^2 = 3 - 2\cos{i} \sqrt{\bar{a}(1 - e^2)} - \frac{1}{\bar{a}},
\label{eq:tiss9}
\end{equation}
which, in combination with Eq.~\ref{eq:tiss5} gives (see also \cite{Strange2009})
\begin{equation}
\bar{C}_T = 3 - \bar{v}_{\infty}^2.
\label{CT_vinfty}
\end{equation}
According to Eq.~\ref{eq:tiss9}, the semi-major axis, eccentricity and inclination determine the magnitude of the hyperbolic excess velocity, and the same value of $v_{\infty}$ can be obtained with different combinations of these parameters. 

\section{The Tisserand Graph}
\label{sec:TG}
\begin{figure}[htb]
\centering
\includegraphics[width=4.0in]{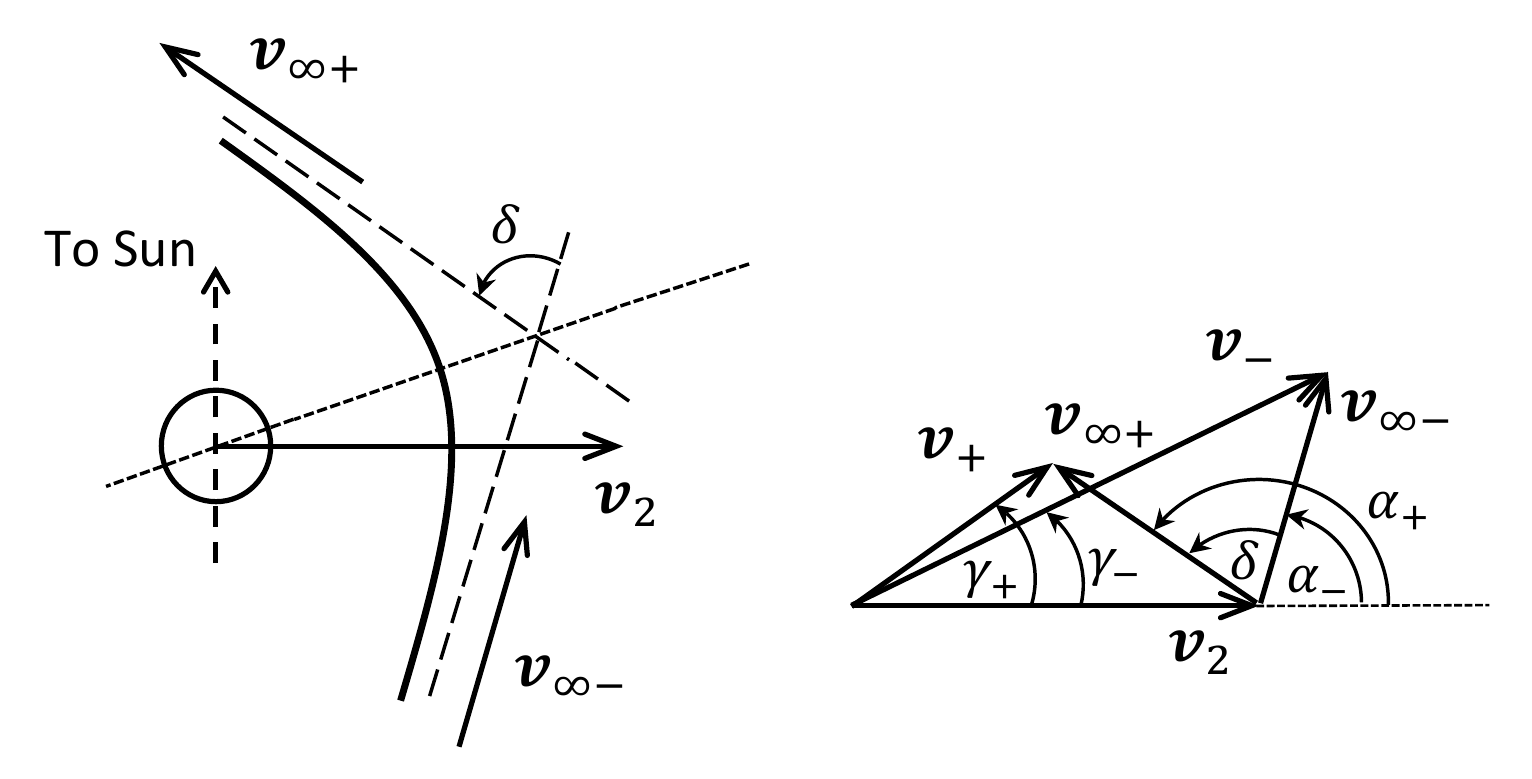}
\caption{Geometry of a flyby (left) and the associated velocity vector diagram (right).}
\label{fig:FlyBy_Geometry_Triangle}
\end{figure}

In the planar approximation (i.e., $i = 0$), $C_T$ depends only on two orbital elements, semi-major axis and eccentricity.  
Figure~\ref{fig:FlyBy_Geometry_Triangle} shows the geometry of a flyby. The time duration of the event is assumed negligible with respect to  
the orbital period of the planet, hence the heliocentric position of the S/C through the flyby is approximately constant. As a result of the close passage, the velocity of the S/C relative to the planet changes. The net effect is a rotation of the inbound hyperbolic excess velocity ${\bf v}_{\infty-}$ by an angle $\delta$, yielding the outbound hyperbolic excess velocity ${\bf v}_{\infty+}$ (Fig.~\ref{fig:FlyBy_Geometry_Triangle}). Hereinfter, the magnitude of these two vectors will be indicated with $v_\infty$. The heliocentric velocity of the S/C changes from  ${\bf v}_-$ to  ${\bf v}_+$, and this change affects both the direction and the magnitude of the vector. The pump angle $\alpha_-$ (respectively, $\alpha_+$) is defined as the angle between ${\bf v}_2$ and ${\bf v}_{\infty-}$ (respectively, ${\bf v}_2$ and ${\bf v}_{\infty+}$). We shall refer to  $\alpha_-$ as the entry pump angle and to $\alpha_+$ as the exit pump angle. Due to 
the symmetry of the problem, we shall limit the discussion to pump angles in the range $[0^{\circ},180^{\circ}$]\footnote{Switching the sign of the pump angle has no effect on the S/C heliocentric trajectory. It simply changes the sign of the radial S/C velocity while leaving the circumferential component unchanged. Thus, the two signs correspond to a pair of symmetric points of the same conic section.}.
From geometry,
\begin{equation}
\alpha_+ = \alpha_- + \delta,
\label{eq:alpha}
\end{equation}
and $\delta$ is $\in$ $[0^{\circ},180^{\circ}$].

\begin{figure}[htb]
  \centering
  \includegraphics[width=7.0in]{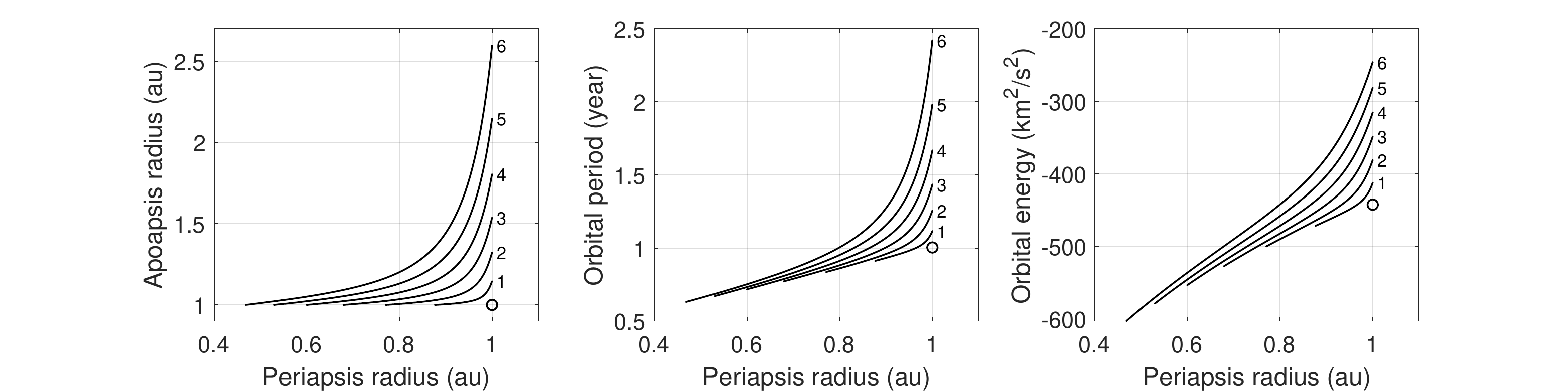}
  \caption{Three versions of the same TG for different choices of the orbital parameters: apoapsis radius (left), orbital period (center) and orbital energy (right) versus periapsis radius. The circle represents the flyby planet, the Earth in this case, and the $v_{\infty}$ contours correspond to values of the hyperbolic excess speed of 1, 2, 3, 4, 5 and 6 km/s, respectively.}
  \label{fig:Tisserand_Earth_Multi}
\end{figure}

\begin{figure}[htb]
  \centering
  \includegraphics[width=3.0in]{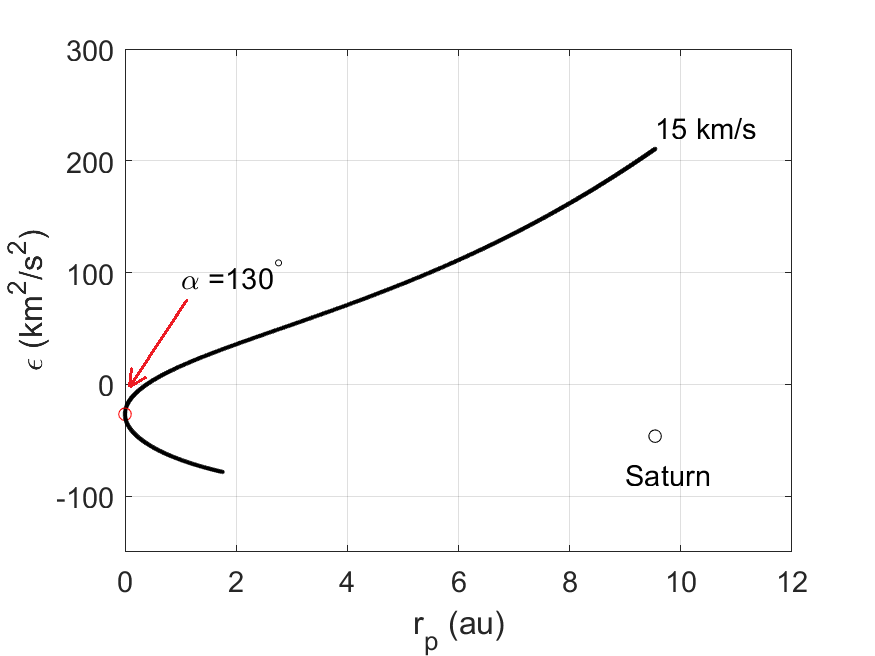}
  \caption{A non-monotonic $v_{\infty}$ contour for Saturn in which retrograde orbits (lower branch of the curve) 
appear for $\alpha > 130$ degrees.}
  \label{fig:TG_retro}
\end{figure}

The law of cosines allows to compute the magnitude of ${\bf v}_-$ (respectively, ${\bf v}_+$) from ${\bf v}_{\infty-}$ (respectively, ${\bf v}_{\infty+}$) and  $\alpha_-$ (respectively, $\alpha_+$):
\begin{eqnarray}
{v}_-^2 & = & v_2^2 + {v}_{\infty}^2 + 2 {v}_{\infty} \cos{\alpha}_-, \label{eq:v3-}\\
{v}_+^2 & = & v_2^2 + {v}_{\infty}^2 + 2 {v}_{\infty} \cos{\alpha}_+= v_2^2 + {v}_{\infty}^2 + 2 {v}_{\infty} \cos({\alpha}_-+\delta). \label{eq:v3+}
\end{eqnarray} 
Given ${v}_{\infty}$ and $\alpha$ ($\alpha_-$ or $\alpha_+$), the above formulas yield the magnitude of the heliocentric velocity (${v}_-$ or ${v}_+$). Then, the vis-viva integral (Eq.~\ref{eq:tiss2}) gives the corresponding semimajor axis, and the orbital energy ${\epsilon}$ follows from   
\begin{equation}
\epsilon = -\frac{\mu_1}{2a}.
\label{eq:energy}
\end{equation}
Equation~\ref{eq:tiss9} can then be solved for the eccentricity. In this way, curves of constant $v_\infty$  can be represented in a 2D map whose axes portrait the semimajor axis and the eccentricity, or any equivalent pair of orbital parameters (for example, the orbital energy versus the periapsis radius, the orbital period versus the periapsis radius or, only for elliptical orbits, the apoapsis radius versus the periapsis radius, Fig.~\ref{fig:Tisserand_Earth_Multi}). This 2D map is the TG and is employed to visualize the effect of planetary flybys on the heliocentric Keplerian orbits of the S/C.
Since ${\bf v}_{\infty-}$ and ${\bf v}_{\infty+}$ have the same magnitude, they correspond to the same $v_\infty$ contour.
Note that in Fig.~\ref{fig:Tisserand_Earth_Multi} the contours $v_\infty$ = cons. are always monotonic, irrespective of the choice of variables. This is a consequence of  Eq.~\ref{eq:tiss9}, which yields a unique value of $e$ for a given ($v_\infty$, $a$) pair. That is, the function $e(a)$ for fixed $v_\infty$  is single-valued. Therefore, as it is continuous, it is also monotonic. However, this implicitly assumes that the orbital inclination is unique, but, when both $v_\infty$ and $\delta$ are very large, retrograde heliocentric orbits appear.  
If both prograde and retrograde orbits in the same plane are considered, there are two corresponding values of the inclination with opposite sign of the cosine.  
Thus, Eq.~\ref{eq:tiss9} gives two eccentricities for each ($v_\infty$, $a$) pair and monotonicity is lost (see Fig.~\ref{fig:TG_retro}). This is undesirable, as it makes finding contour intersections much more complex. Monotonicity can be recovered by partitioning each contour into a prograde and a retrograde branch.

\subsection{Construction of the contours}
\label{sec:construct}
\begin{figure}[htb]
  \centering
  \includegraphics[width=5.0in]{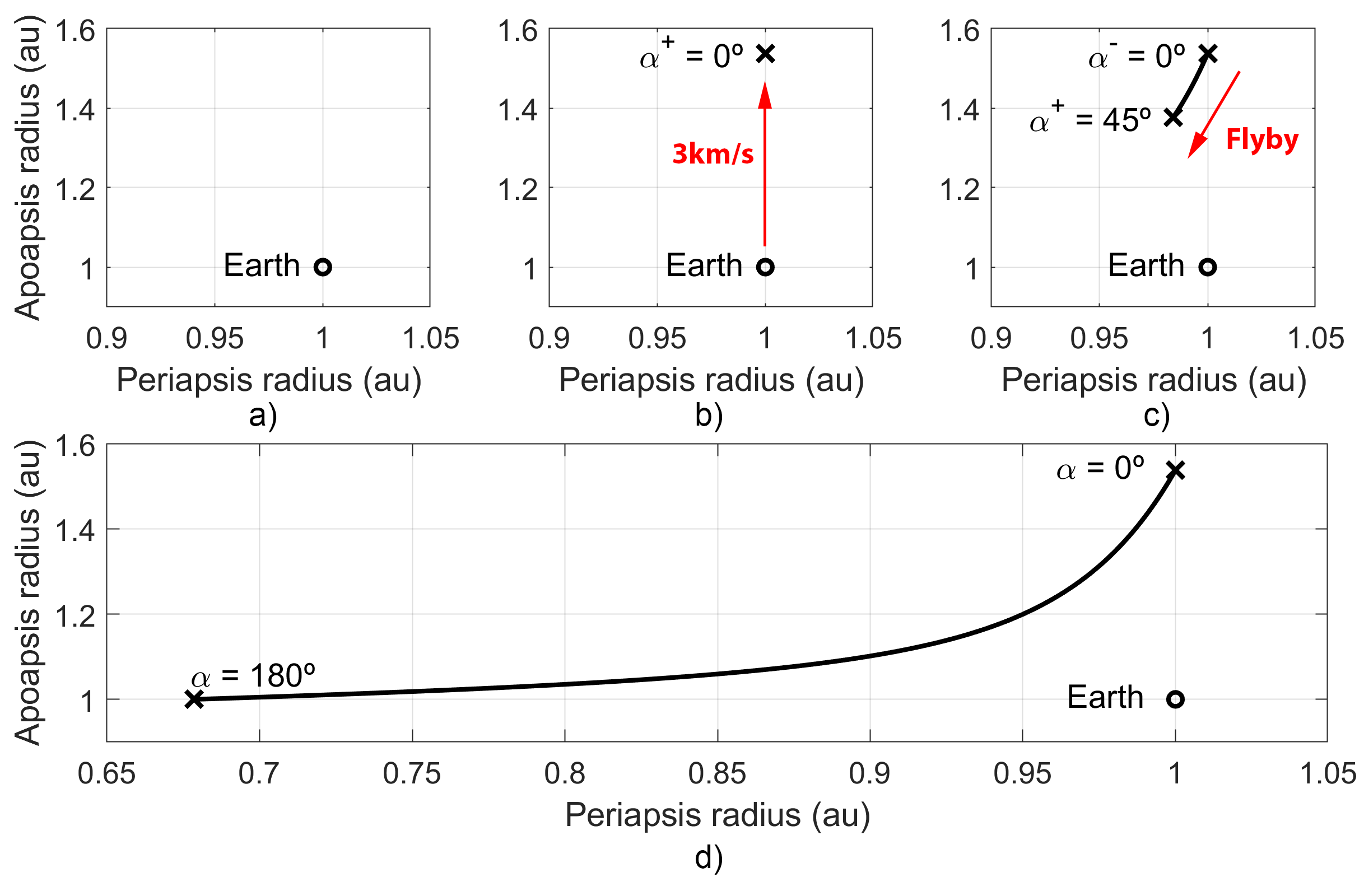}
  \caption{Construction of a $v_{\infty}$ contour.}
  \label{fig:Tisserand_Tutorial}
\end{figure}

The TG of Fig.~\ref{fig:Tisserand_Tutorial} illustrates the steps taken in the construction of a $v_{\infty}$ contour. The case shown corresponds to a $v_{\infty}$ of 3 km/s relative to Earth. The circular orbit of the planet is the point (1 au, 1 au) (Fig.~\ref{fig:Tisserand_Tutorial}a)\footnote{Note that the periapsis of the heliocentric orbit of the S/C can never be higher than the orbit of the flyby planet.}. The upper point of the contour is obtained by setting $\alpha$ = 0. The algorithm outlined above gives a periapsis radius of 1 au (in this case, the S/C's orbit is tangent to Earth's orbit) and an apoapsis radius of 1.54 au (Fig.~\ref{fig:Tisserand_Tutorial}b). 
Varying $\alpha$ between 0 and 180 degrees yields the entire contour (Fig.~\ref{fig:Tisserand_Tutorial}d).

When $v_{\infty}$ contours of different planets are plotted in the same TG, their intersections correspond to orbits that can be linked by flybys with these planets. 
For example, in the sequence of Fig.~\ref{fig:Strange1}, the S/C departs Earth on a 3 km/s contour, performs a flyby with Venus at 5 km/s of $v_{\infty}$ and passes by the Earth with a relative speed of 9 km/s twice (resonant flybys) before reaching Jupiter at a relative speed of 6 km/s.

A $v_{\infty}$ contour can also be used to construct sequences of consecutive flybys with the same planet and characterized by the same hyperbolic excess speed. For example, in Fig.~\ref{fig:Tisserand_Tutorial}c, after tangential departure from Earth, the S/C returns after an integer number of revolutions, hence with the same encounter geometry ($\alpha_-$ = 0) and heliocentric velocity. If $\alpha_+$ = 45 degrees, the new periapsis radius is 0.98 au and the apoapsis radius is 1.37 au.   

Due to the relationship (see, e.g., \cite{Curtis2014} Chapt. 2, pp. 100-101) 
\begin{equation}
\displaystyle \sin \left(\frac{\delta}{2}\right) = \left(1+\frac{r_{\pi} v_{\infty}^2}{\mu_2}\right)^{-1}
\label{eq:rpi}
\end{equation}
between  $\delta$ and the pericenter radius $r_{\pi}$ of a hyperbola with a given $v_{\infty}$ and focus at $\mu_2$, imposing a minimum flyby height sets an upper limit $\delta_{max}$ to the achievable deflection angle. For example, if the minimum periapsis height above the surface of the Earth on a $v_{\infty}$ contour of 3 km/s is 200 km, $\delta_{max}$ = 121 degrees. $\delta_{max}$ in turn limits the maximum displacement along a $v_{\infty}$ contour that can be achieved with a single flyby.

\subsection{Selection of the $v_{\infty}$ levels}
\label{sec:vinfty}
\begin{figure}[h!]
\centering
\includegraphics[width=5.0in]{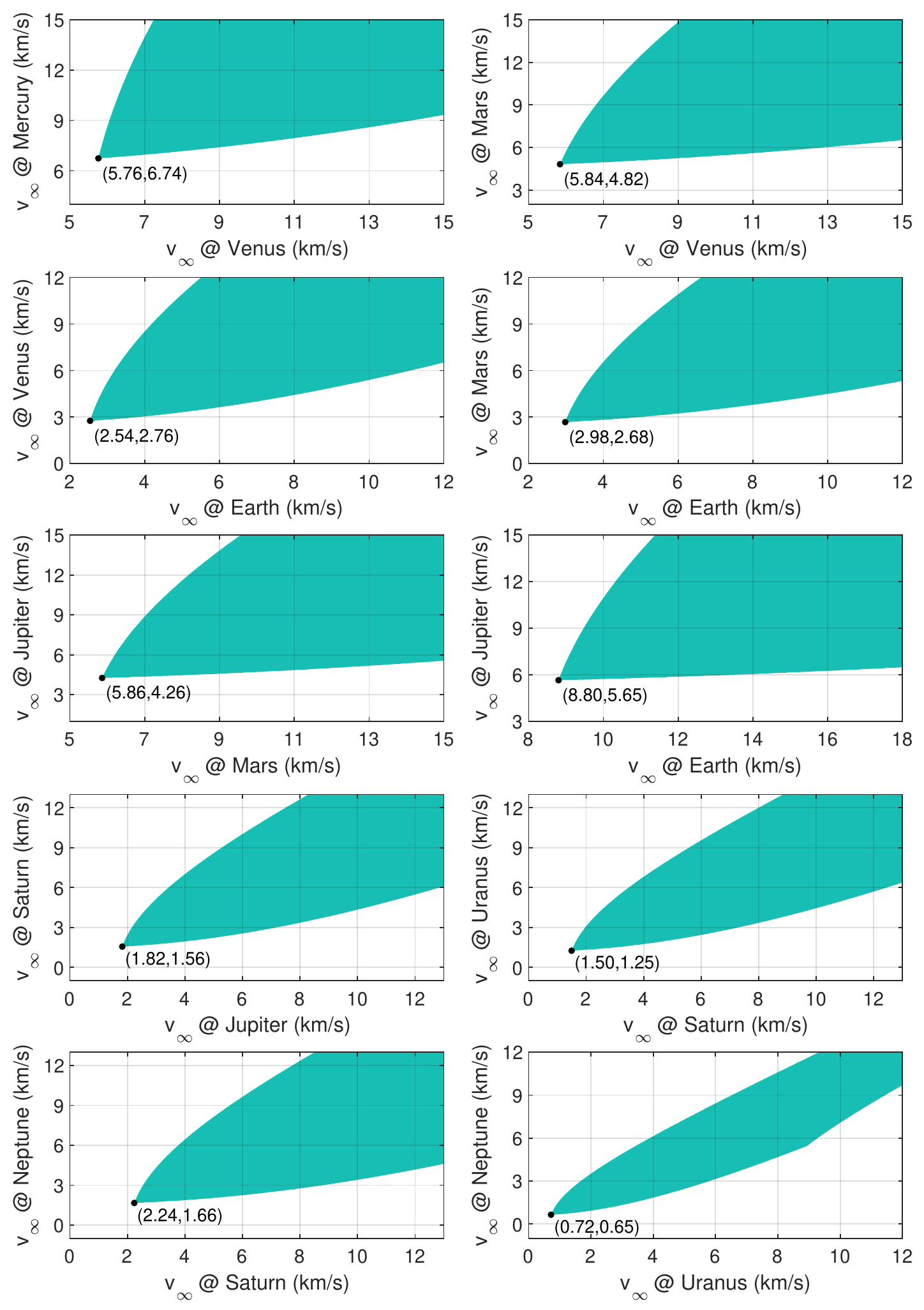}
\caption{Maps of intersections between $v_{\infty}$ contours of different planets.}
\label{fig:vinf_choice}
\end{figure}

Constructing a TG requires the selection of the planets and the identification of suitable $v_{\infty}$ levels for each of them. Even for the experienced orbit analyst, this task is not trivial, especially when the 
solution space is wide and involves several bodies. Furthermore, since the goal is to build sequences of flybys from departure to arrival, only those pairs of $v_{\infty}$ contours that intersect each other are useful. Figure~\ref{fig:vinf_choice} provides some insight into this problem. Each panel refers to a specific pair of planets. The axes 
report $v_{\infty}$ values for each planet, so that each point in the diagram can be associated with a pair of $v_{\infty}$ contours. The shaded area shows the combinations of $v_{\infty}$. The apex (marked with a solid circle) corresponds to a Hohmann transfer between the two planets. It is the heliocentric ellipse that intersects both circular orbits with minimum hyperbolic excess velocity at departure and arrival. Therefore, the apex must lie at the bottom left corner of the shaded region.
The charts can be used to narrow quickly the range of $v_{\infty}$ contours that must be explored. For example, consider an Earth-Mars transfer for which the combination of launcher and payload limits the departure $C_3$ to 36 km$^2$/s$^2$ ($v_{\infty}$ = 6 km/s). From Fig.~\ref{fig:vinf_choice}, the useful $v_{\infty}$ values for Earth lie between 3 and 6 km/s, while in the case of Mars we have to include contours from 2.7 to 10.5 km/s. Points (i.e., pairs of $v_{\infty}$ values) outside the shaded area can safely be ignored, as they cannot yield an intersection. That is the case, for example, for the combination of $v_{\infty}$ = 7 km/s at Mars and 11 km/s at Jupiter. This can be crosschecked against Fig.~\ref{fig:Strange2}, which shows that, as expected, there is no intersection between those contours. 

\section{Tisserand PathFinder}
\label{sec:tpf}

\begin{figure}[htb]
  \centering
  \includegraphics[width=5.0in]{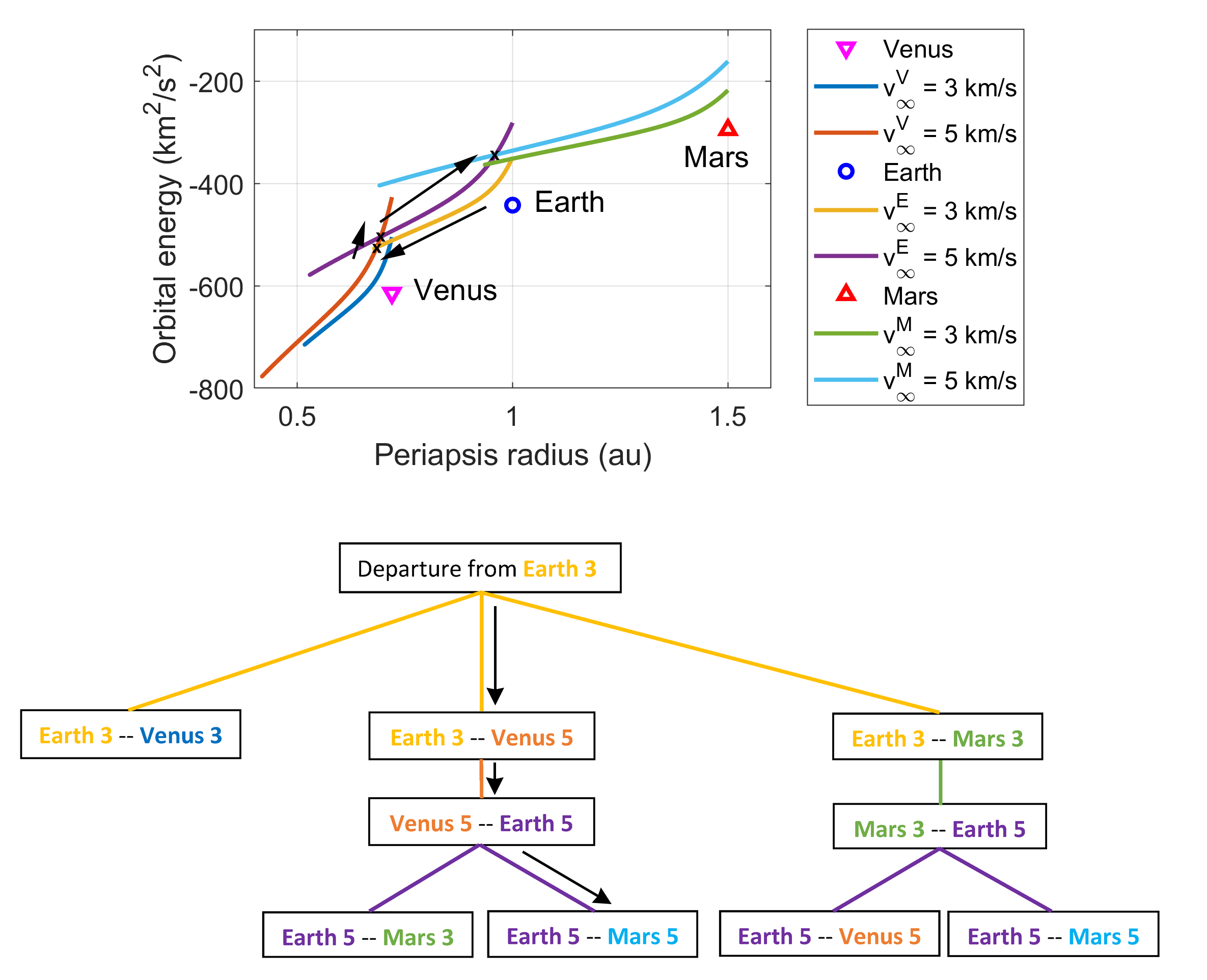}
  \caption{TG (top) and the corresponding tree of encounter paths (bottom) for an Earth-to-Mars transfer in which  flybys with Earth, Venus and Mars are considered. The arrows indicate an Earth-Venus-Earth-Mars path.}
  \label{fig:Tisserand_Graph_vs_Tree}
\end{figure}

The TPF algorithm is based on the representation of the TG as a tree structure: 
an intersection between contours in the graph is a tree node and a transfer between
intersections along a contour is a branch. The set of nodes and branches constitutes the tree.
Each node has one parent node and may have one or more children nodes. A tree search algorithm has been applied to this representation in order to traverse the graph in an ordered way, collect transfers between planets and form encounter paths. Figure~\ref{fig:Tisserand_Graph_vs_Tree} (top) illustrates an interplanetary TG for a transfer from Earth to Mars with Earth, Venus and Mars flybys and two $v_{\infty}$ contours for each planet (3 and 5 km/s). 
The black crosses mark contour intersections and the arrows signal transfers between planets.
The construction of the tree structure from the TG is shown in the bottom part of the figure: starting from Earth 
with $\alpha$ = 0 and $v_{\infty}$ = 3km/s (root node), the possible paths are determined traversing the nodes and branches of the tree. The path indicated by the arrows goes through four nodes (Earth 3, Earth 3 - Venus 5,  Venus 5 -  Earth 5, Earth 5 - Mars 5) and three branches (Earth 3 to Venus 5, Venus 5 to Earth 5, Earth 5 to Mars 5).

\subsection{Determination of the intersection of two $v_{\infty}$ contours}
\label{sec:inters}
\begin{figure}[h!]
\centering
\includegraphics[width=2.6in]{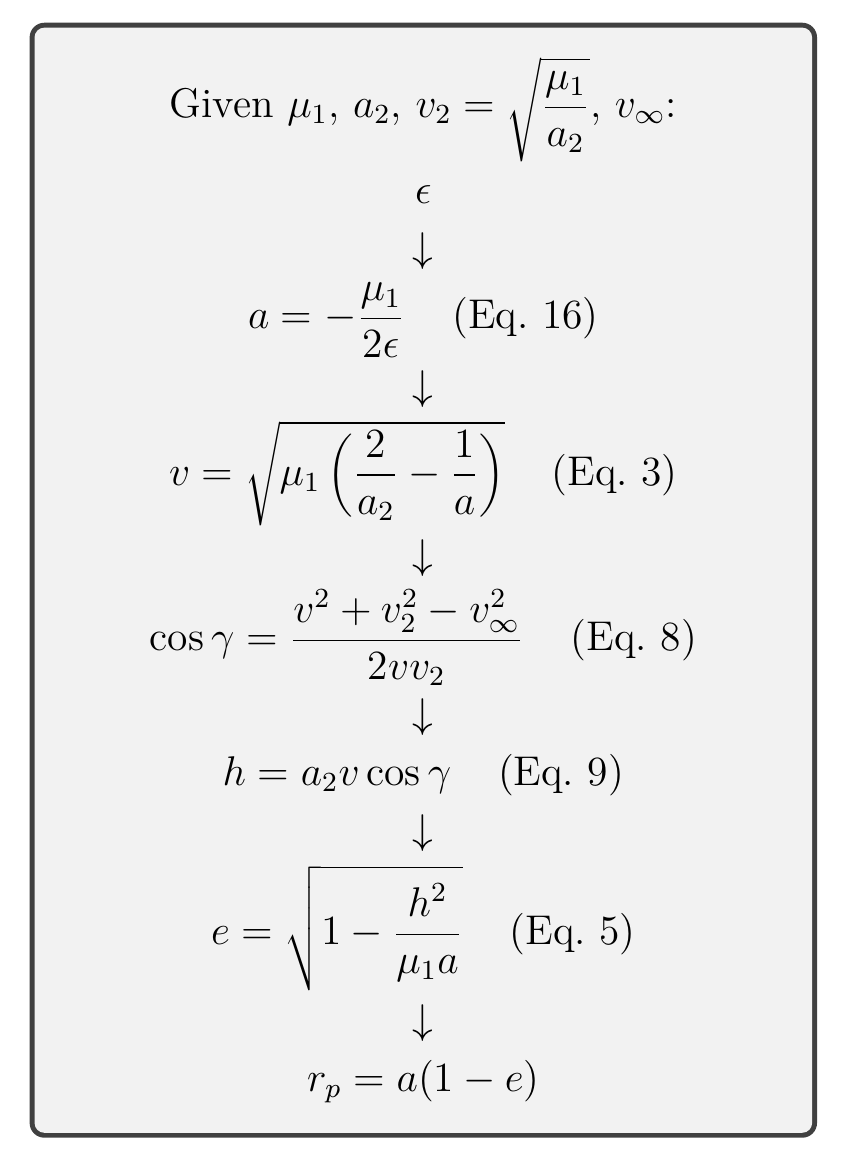}
\caption{Algorithm to obtain $r_{p}$ from $\epsilon$ in a $v_{\infty}$ contour relative to a planet with orbital radius $a_2$.}
\label{fig:eps_rpi}
\end{figure}

\begin{figure}[htb]
  \centering
  \includegraphics[width=3.5in]{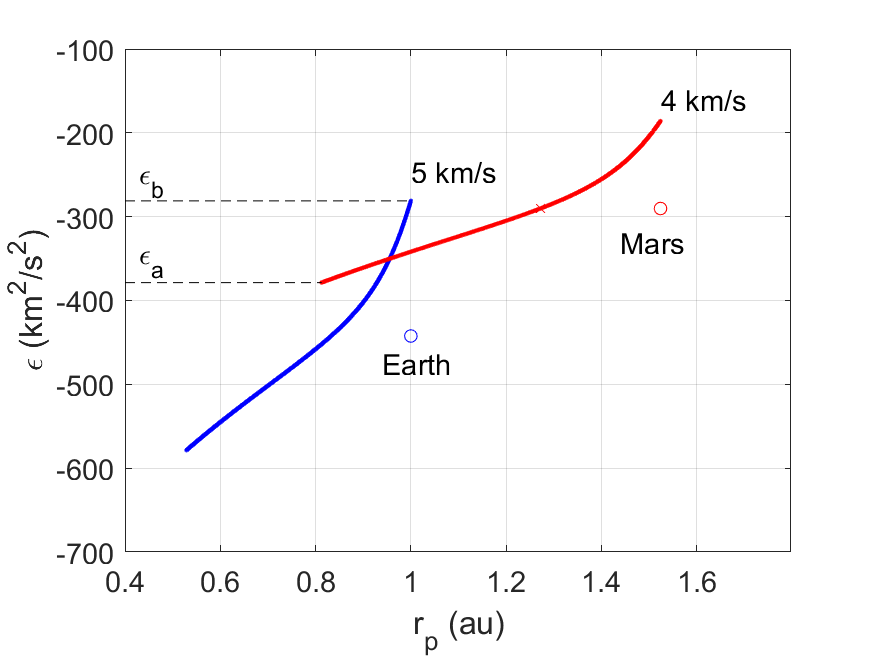}
  \caption{Two intersecting contours relative to Earth and Mars with $v_{\infty}$ of 5 and 4 km/s, respectively.}
  \label{fig:TG_intersection}
\end{figure}

The intersections between $v_{\infty}$ contours are determined finding the zeros of the function
\begin{equation}
f(\epsilon) = r_{p2}(\epsilon) - r_{p1}(\epsilon),
\label{eq:intersection}
\end{equation}
where the orbital energy is taken as the independent variable and $r_{p1}$ and $r_{p2}$ are the periapsis radii on the two contours.
To solve Eq.~\ref{eq:intersection}, the regula-falsi technique \cite{Dahlquist2003} has been chosen due to its simplicity (no derivatives required) and robustness. The Illinois variant \cite{Dowell1971} of the algorithm is used for improved performance.
At every iteration, $\epsilon$ is used to determine $r_{p}$ in both contours following the procedure outlined in Fig.~\ref{fig:eps_rpi}. 
The initial search interval is the range of $\epsilon$ common to both contours, i.e., between 
$\epsilon_a = max(min(\epsilon_1),min(\epsilon_2))$ and $\epsilon_b=min(\epsilon_1(\alpha=0^{\circ}),\epsilon_2(\alpha=0^{\circ}))$, as shown in the example of Fig.~\ref{fig:TG_intersection}.
The iterations stop when the absolute value of $f(\epsilon)$ falls below a specified tolerance. As a reference, it takes up to six iterations to reach an accuracy of 10 km.

\subsection{Tree search}
Tree search algorithms \cite{Cormen1990} are often used in optimization problems in which, given a starting point, several options must be explored to find the optimal solution. In our case, the goal is building a comprehensive set of candidate solutions, i.e., to find all the paths in the TG connecting the departure planet with the target planet, regardless of their performance. Driven by this requirement, we chose an uninformed depth-first algorithm \cite{Korf1985}: the uninformed tree search is suitable when there is no {\it a priori} knowledge of the tree, and the depth-first variant (which initially explores the nodes at the deepest levels of the tree and backtracks when it hits a dead-end) offers high speed and low memory consumption when dealing with complex trees.

The main drawback of the depth-first method is that it can get trapped in loops (in our case, repeated flybys with the same planet). To limit the number of planetary encounters and the computation time, the maximum depth of the tree is set by the user. Once the maximum depth is reached, the search does not proceed further along the current branch of the tree. This modified version of the depth-first method is referred to as the depth-limited search method.
Program data is arranged in two main structures:
\begin{itemize}
	\item Node database: It contains a list of the active nodes. Each node has a unique identifier (ID) and is associated with a parent node ID, a planet and specific values of $v_{\infty}$ and $\epsilon$. The parent node is the preceding encounter in the sequence of flybys. Once a node reaches the target destination, the parent IDs are used to rebuild the sequence of encounters.
	\item Stack: This is a FIFO (First-In First-Out) heap where the identifiers of nodes in process are stored. The depth-first search method uses the FIFO strategy: as the tree is expanded vertically, the most recently created nodes (i.e., those at the lowest levels of the tree) are processed first.
\end{itemize}

\begin{figure}[htb]
	\centering
	\includegraphics[width=4.0in]{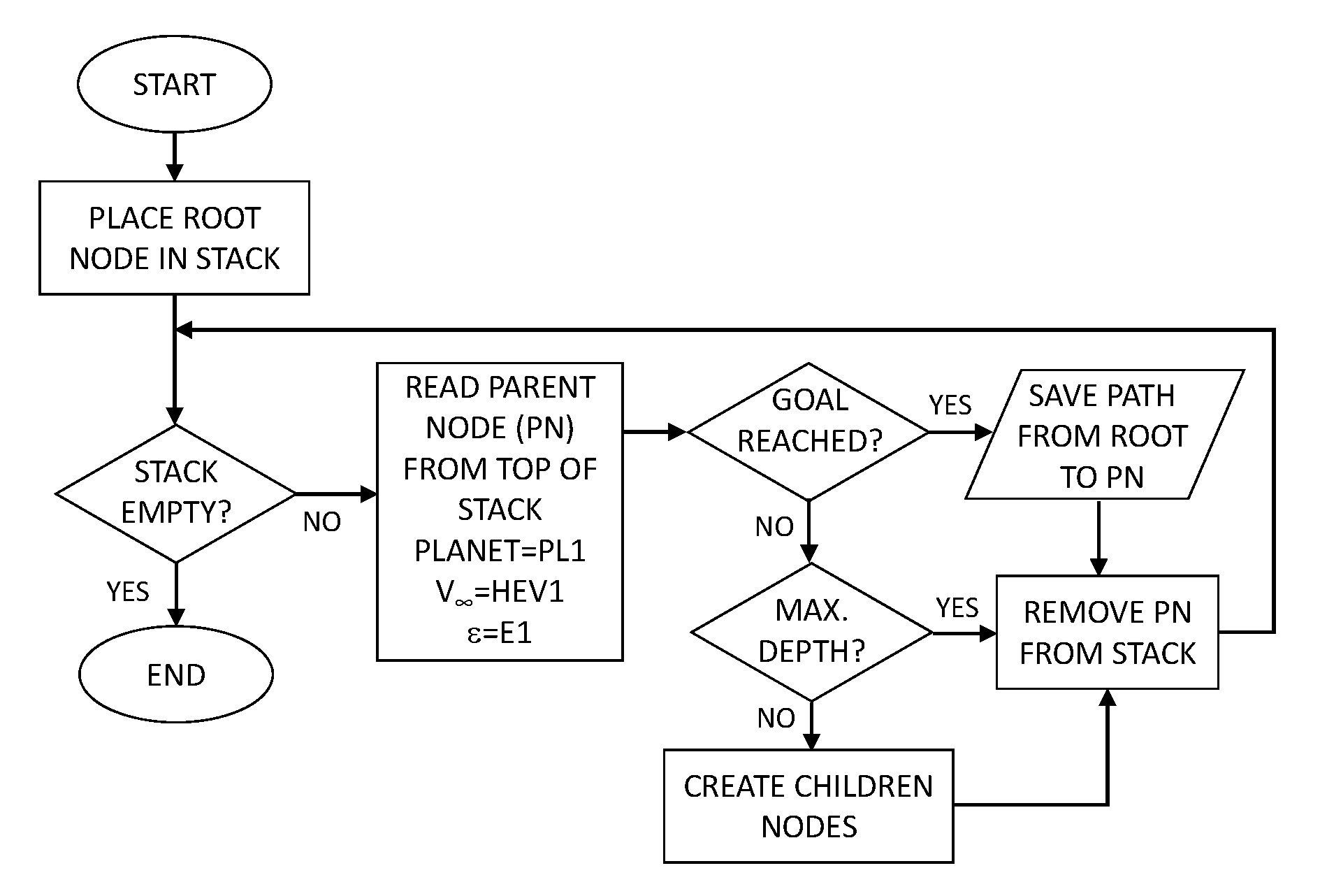}
	\caption{Global functional diagram of TPF.}
	\label{fig:TPF_global}
\end{figure}

\begin{figure}[htb]
	\centering
	\includegraphics[width=5.0in]{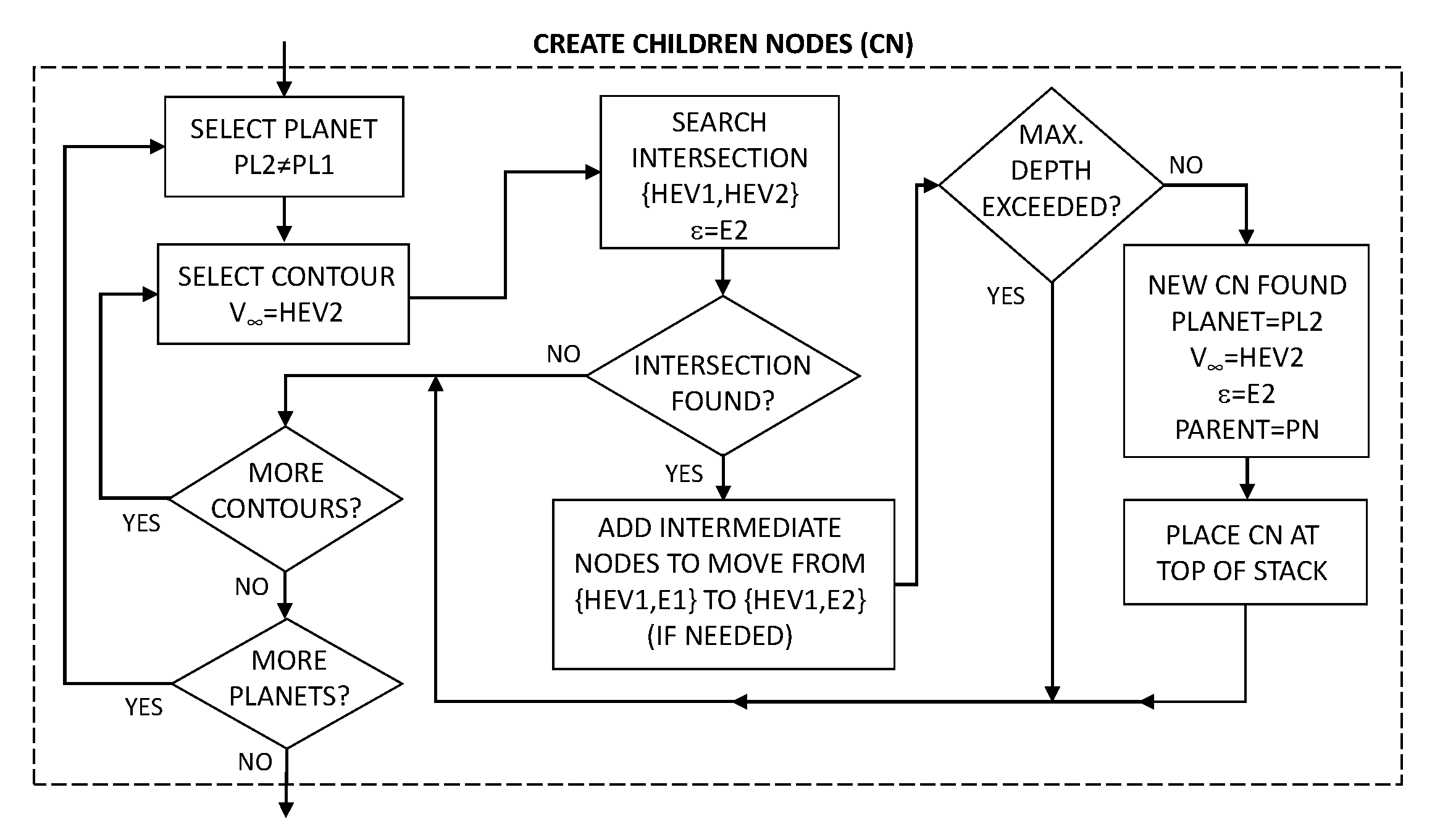}
	\caption{Flowchart of the children node generation block.}
	\label{fig:TPF_children}
\end{figure}

\subsection{Search algorithm}
To make the explanation easier to follow, we shall split the program structure in two distinct functional blocks:
\begin{itemize}
	\item Processing of parent nodes (Fig.~\ref{fig:TPF_global}): These nodes are stored in the stack and were in turn created from other parents in previous iterations of the algorithm.
	\item Generation of children nodes (Fig.~\ref{fig:TPF_children}): These are the nodes that can be reached from the parent nodes by means of a flyby, and subsequently become new parents.
\end{itemize}

\subsubsection{Processing of parent nodes}
The nodes on the stack are examined and their children are generated through these steps:
\begin{enumerate}
\item Read input data: List of planets and $v_{\infty}$ contours associated with each one, departure conditions (planet and $v_{\infty}$ corresponding to the launch energy) and target (arrival planet and, if applicable, range of $v_{\infty}$).	
\item Initialize the stack with the root node (departure planet and $v_{\infty}$).
\item If the stack is empty, terminate the program. Otherwise, read the node at the top of the stack. It becomes the current parent node (PN), associated with planet PL1, $v_{\infty}$=HEV1 and specific orbital energy $\epsilon$=E1.
	\item Check if PN has reached the final state specified by the user (i.e., target planet and $v_{\infty}$ level). If that is the case, rebuild the path from departure to destination planet using the parent node IDs, save it and proceed to step 6.
	\item If PN has not reached the user-defined maximum depth, find its children nodes (see below).
	\item Purge PN from the stack and go back to step 3.
\end{enumerate}

\subsubsection{Generation of children nodes}
The sequence of steps to create new Children Nodes (CN) is:
\begin{enumerate}
	\item Select a candidate planet PL2 to explore (PL2$\ne$PL1).
	\item Select one of the $v_{\infty}$ levels for PL2 (HEV2).
	\item Search for an intersection between contours HEV1 of PL1 and HEV2 of PL2. Let E2 be the orbital energy of the intersection point (if it exists). If no intersection is found, proceed to step 6.
	\item Add as many intermediate nodes as needed to move from E1 to E2 along the HEV1 contour. These nodes represent consecutive flybys with PL1 to achieve sufficient deflection angle without colliding with the planet. The intermediate nodes give rise to linear branches of the tree (i.e., no bifurcations) and they are never placed in the stack (because they require no further analysis). Intermediate nodes are important, however, because they add to the tree depth.
	\item If the maximum depth has not been exceeded, the intersection becomes a CN and is placed at the top of the stack. PL2, HEV2, E2 and PN are stored in the database entry for this CN.
	\item If there are more $v_{\infty}$ levels to explore for PL2, go back to step 2.
	\item If there more planets to process, go back to step 1.
\end{enumerate}

\subsection{Garbage collection}
Memory consumption can be an issue when evaluating large tree structures. To reduce the memory footprint of TPF, nodes that are no longer useful (e.g., nodes with no children or nodes from a branch that has been entirely explored) are deleted from the database (pruned) periodically. Because the database does not contain all the nodes analyzed, the information required to reconstruct the sequence of encounters is stored in the output file. Whenever a path to the target planet is found, the list of encounters and the corresponding nodes are stored in the output file. The list of nodes in the output data, while only a small subset of those explored, is sufficient to interpret the paths.

A Matlab script implementing the TPF algorithm has been published under LGPL license \cite{DelaTorre2020}.
The results presented in the following sections have been computed in Matlab R2019a running under Windows 10 1803 in an Intel Core i7-6700K CPU with 4.00GHz and 32GB of RAM.
In the discussion, the encounter paths are grouped into planet sequences, i.e., paths that connect the same ordered list of planets regardless of the $v_{\infty}$ levels.

\section{Validation}
\label{sec:test}
\begin{figure}[htb]
\centering
\includegraphics[width=5.0in]{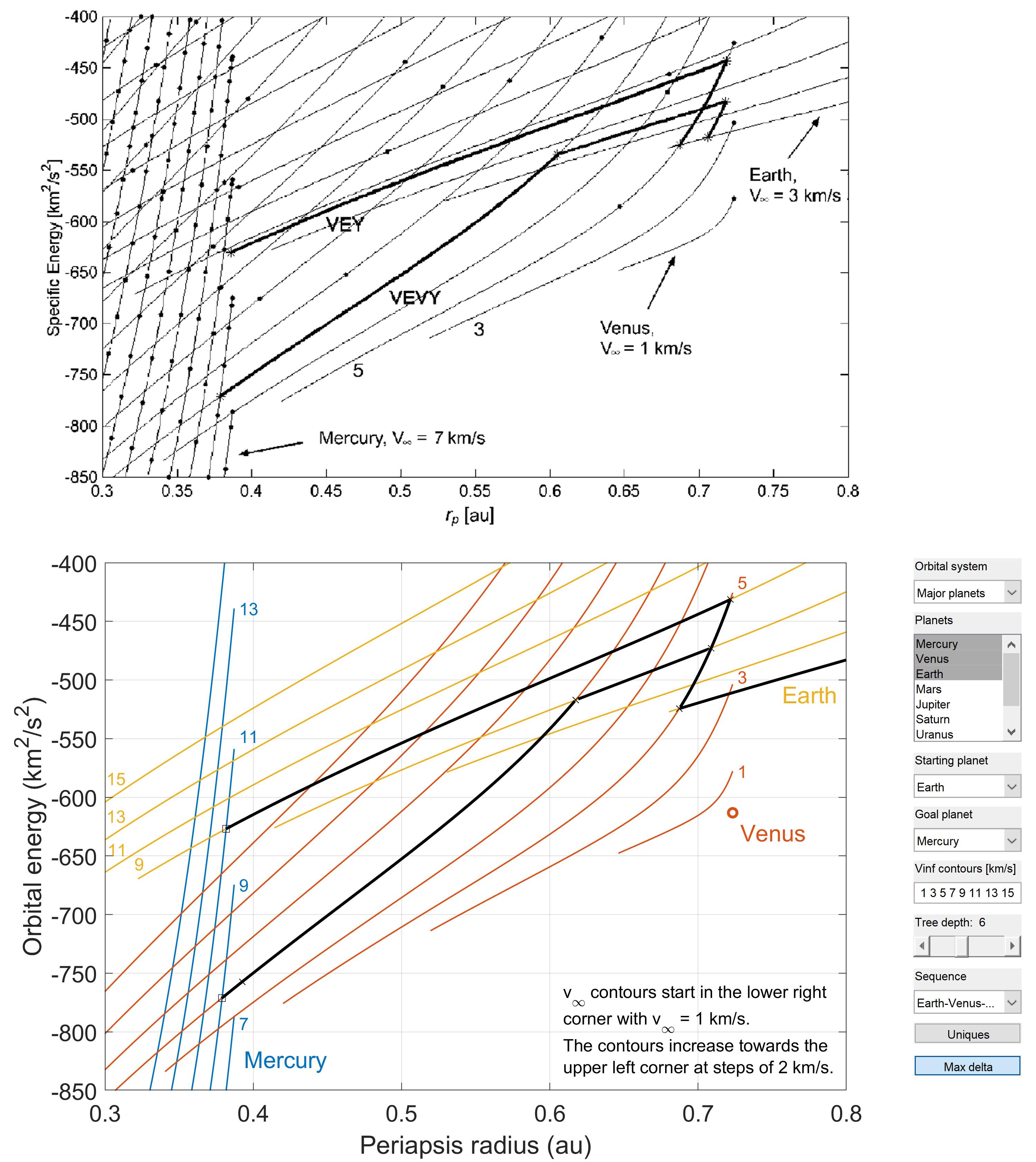}
\caption{Validation case V1: two Earth-to-Mercury paths from \cite{Strange2002} (top) and the same sequences obtained with TPF (bottom). The letters V, E and Y mean Venus, Earth and Mercury, respectively.}
\label{fig:Tisserand_Strange2002}
\end{figure}

\begin{figure}[htb]
\centering
\includegraphics[width=5.0in]{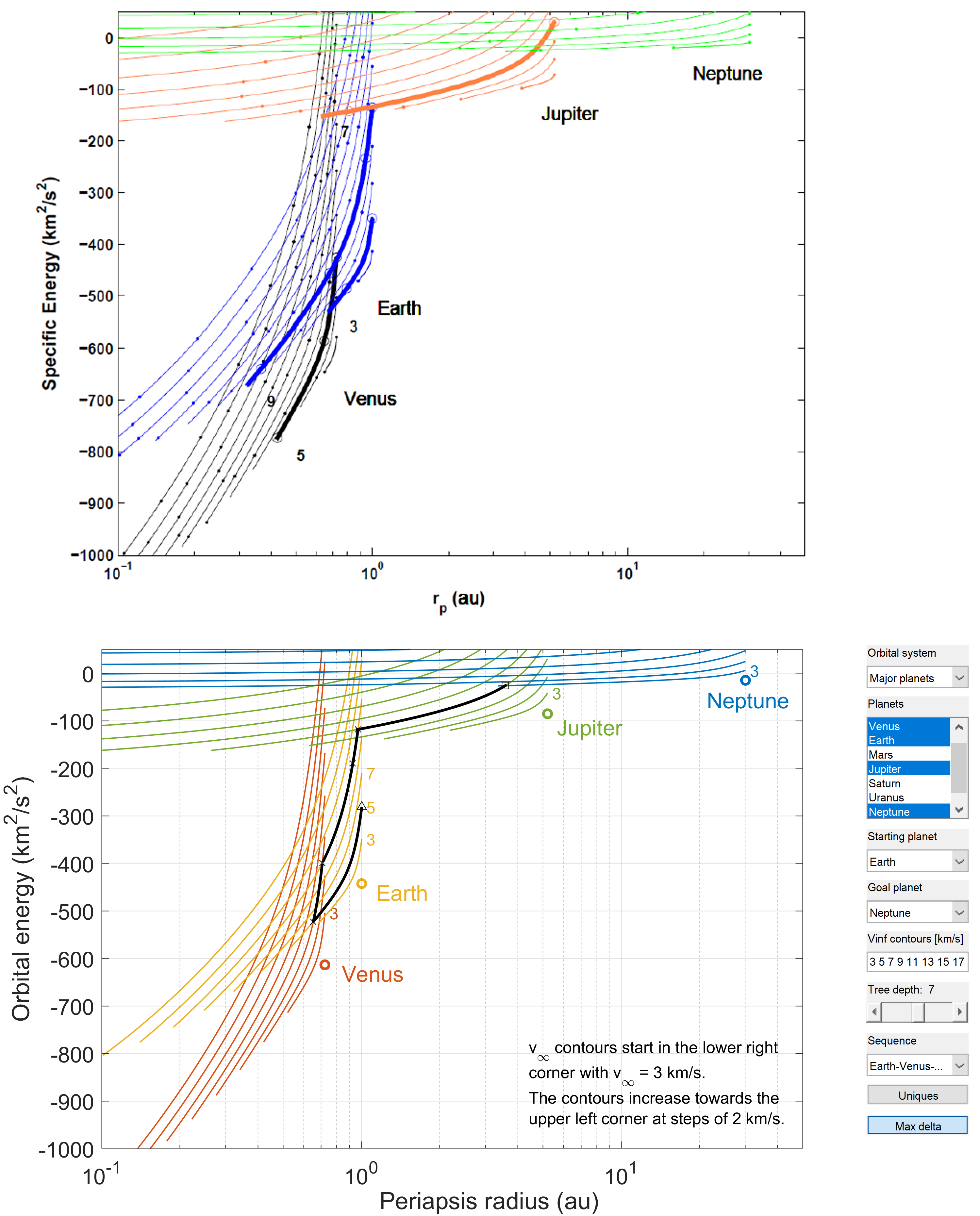}
\caption{Validation case V2: match between the Earth-to-Neptune path of \cite{Hughes2013} (top) and that computed by TPF (bottom).}
\label{fig:Tisserand_Hughes2013}
\end{figure}

\begin{figure}[htb]
\centering
\includegraphics[width=5.0in]{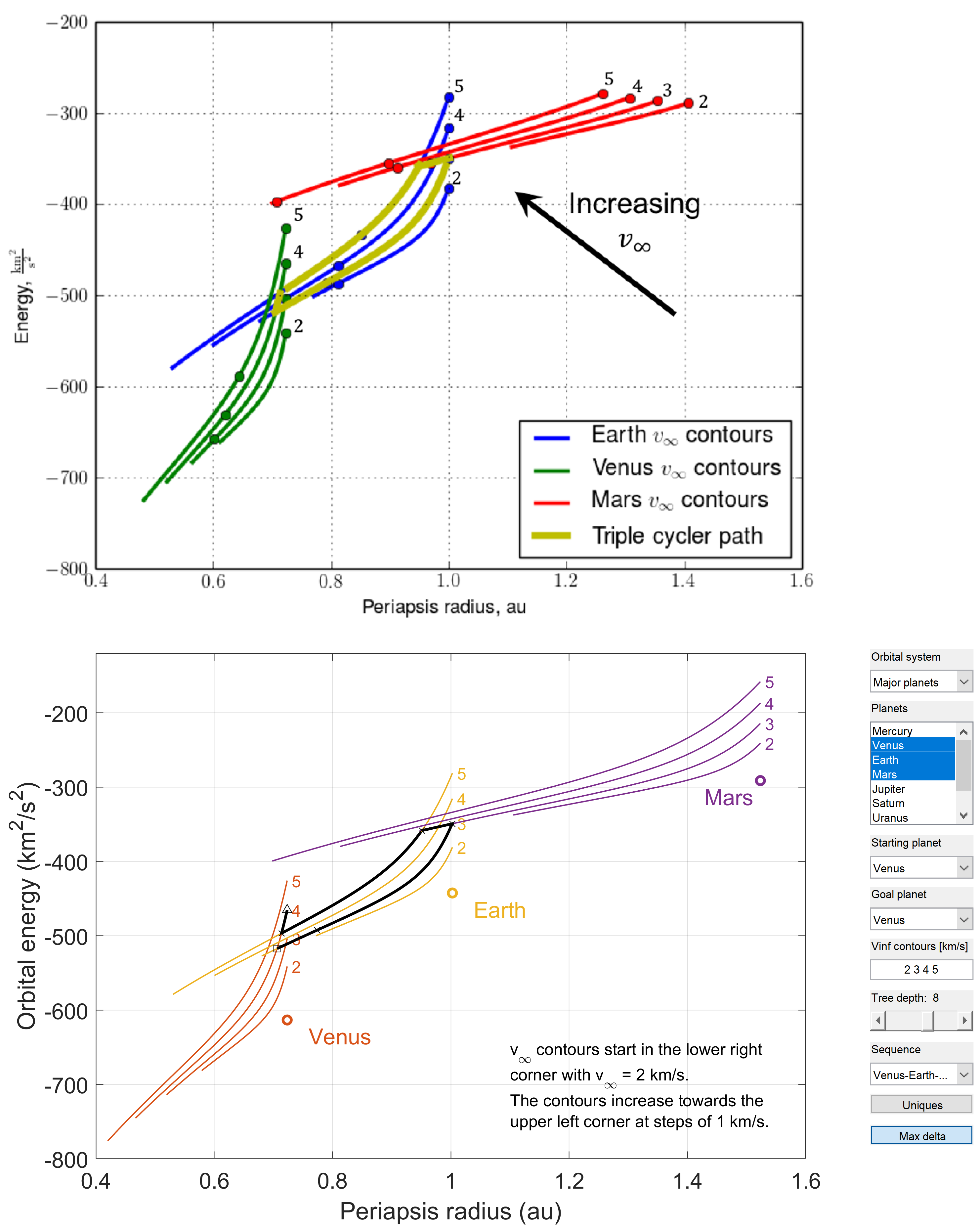}
\caption{Validation case V3: match between the Venus-Earth-Mars cycler path of \cite{Jones2017} (top) and that computed by TPF (bottom).}
\label{fig:Tisserand_Jones2017}
\end{figure}

The results obtained with TPF have been compared with three solutions (named V1, V2, V3) presented in the literature. 
\begin{itemize}
\item V1: from Earth to Mercury. The TG contained in the upper part of Fig.~\ref{fig:Tisserand_Strange2002} shows two encounter paths identified by \cite{Strange2002}: 1) Earth 3, Earth 3 - Venus 5, Venus 5 - Earth 7, Earth 7 - Venus 9, Venus 9 - Mercury 9; 2) Earth 3, Earth 3 - Venus 5, Venus 5 - Earth 9, Earth 9 - Mercury 11.  
TPF yields the same two paths (Fig.~\ref{fig:Tisserand_Strange2002} bottom). Additionally, TPF indicates that the sequence of encounters for the first path is Earth-Venus-Earth-Venus-Venus-Mercury because a transfer to Mercury 9 from Earth 7 requires two consecutive Venus flybys to prevent a collision with the planet (due to the deflection angle limitation).
\item V2: from Earth to Neptune. \cite{Hughes2013} identifies 72 planet sequences between Earth and Neptune. The path marked in Fig.~\ref{fig:Tisserand_Hughes2013} top is Earth 5, Earth 5 - Venus 7, Venus 7 - Earth 11, Earth 11 - Jupiter 7, Jupiter 7 - Neptune 3. TPF is able to find the same sequence (Fig.~\ref{fig:Tisserand_Hughes2013}  bottom).
\item V3: Venus-Earth-Mars cycler. \cite{Jones2017} identifies a triple cycler through Venus, Earth and Mars.
The path highlighted in Fig.~\ref{fig:Tisserand_Jones2017} top is Venus 4, Venus 4 - Earth 5, Earth 5 - Mars 3, Mars 3 - Earth 3, Earth 3 - Venus 4. TPF finds the same solution (Fig.~\ref{fig:Tisserand_Jones2017} bottom).
\end{itemize}
Table~\ref{table:TPF_performance} records the total CPU time, the number of encounter paths found and the number of planet sequences identified for the three validation cases. 

\section{Application to mission design}
\label{sec:app}
The following two mission scenarios have been designed and optimized with the aid of TPF: a transfer from Earth to Mars (M1) and the trajectory of JUICE \cite{EuropeanSpaceAgencyESA2014} (M2). These applications illustrate the advantages of coupling TPF with a trajectory optimizer, allowing fully automated mission design and optimization.
Note that in this case, the distinct paths of each planet sequence are not used, since the optimizer accepts an ordered list of planets and works by varying the encounter dates within each sequence. The deflection angle is limited to prevent trajectories intersecting the planet surface. 

\begin{figure}[htb]
\centering
\includegraphics[width=6.0in]{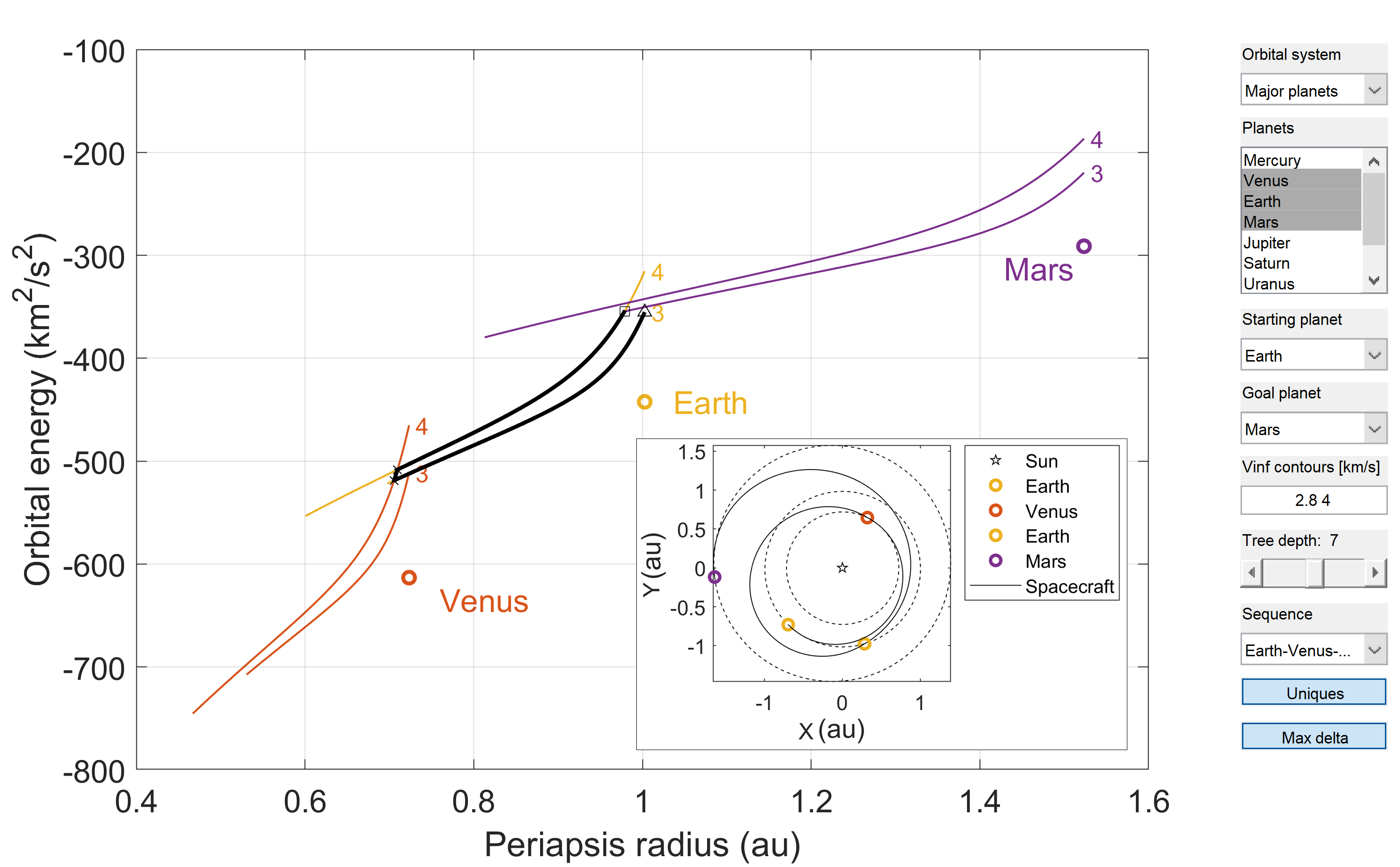}
\caption{Scenario M1: TG for an Earth-to-Mars transfer and the optimized interplanetary trajectory obtained from the sequence highlighted in the diagram.}
\label{fig:M1}
\end{figure}

\begin{figure}[htb]
\centering
\includegraphics[width=6.0in]{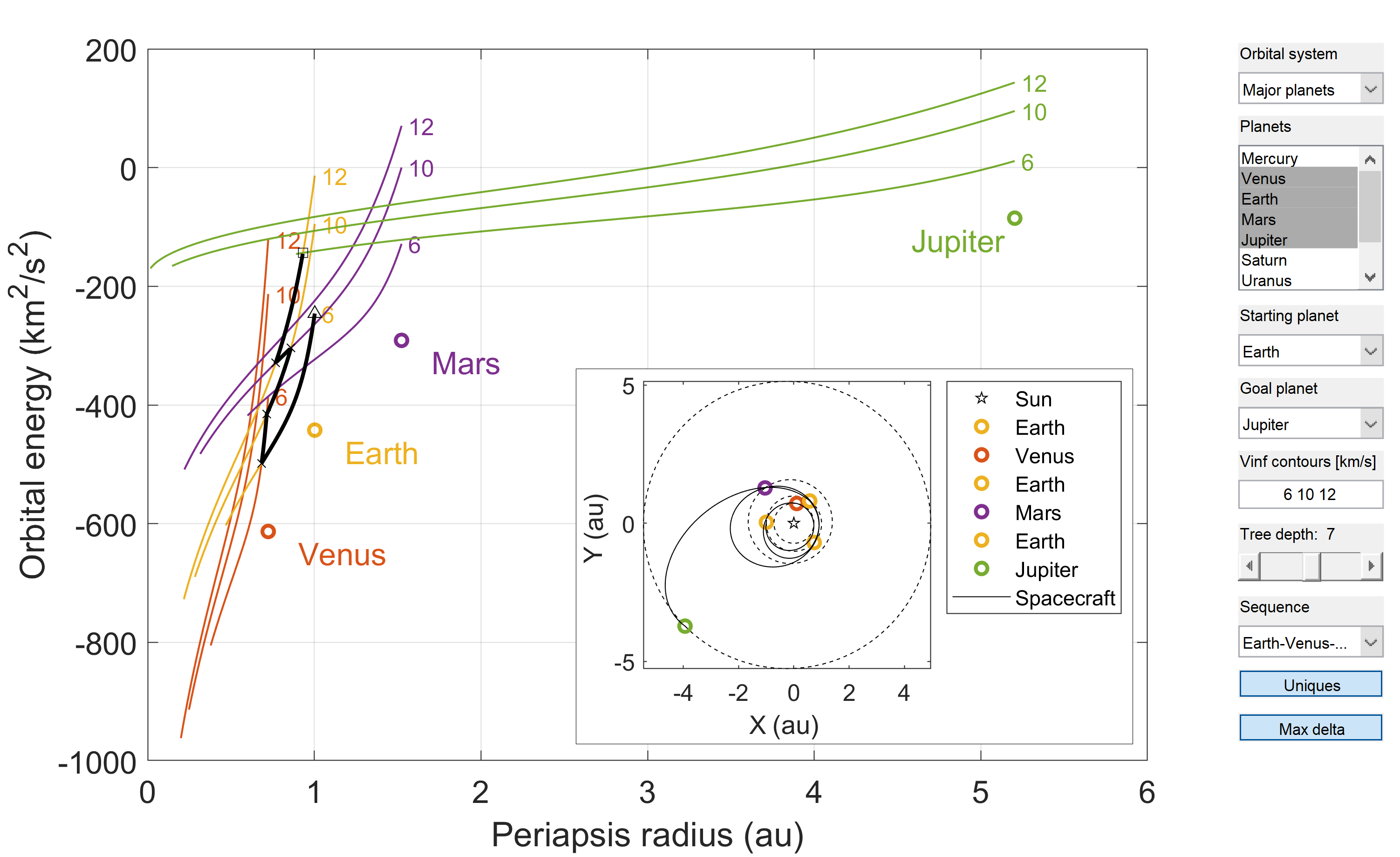}
\caption{Scenario M2: TG for a JUICE-like transfer and optimized trajectory.}
\label{fig:M3}
\end{figure}

The TGs of the two scenarios have a maximum tree depth of seven encounters including the departure and arrival planet. 
\begin{itemize}
\item M1: the TG contains two $v_{\infty}$ contours for each planet (Earth, Venus, Mars). In order to discard direct Earth-to-Mars transfers, the lowest $v_{\infty}$ at Earth is set at 2.8 km/s. The path displayed in Fig.~\ref{fig:M1} is an Earth-Venus-Earth-Mars with Earth 2.8, Earth 2.8 - Venus 4, Venus 4 - Earth 4, Earth 4 - Mars 2.8. From the corresponding planet sequence, the interplanetary trajectory optimizer outputs a solution departing on 07/05/2023 with $v_{\infty}$ of 2.8 km/s and arriving on 19/06/2025 with $v_{\infty}$ of 3.0 km/s. The flybys with Venus and Earth (respectively on 17/10/2023 and 08/70/2024) are powered and require velocity impulses of 913 and 155 m/s, respectively. The departure dates explored range from 01/01/2020 to 01/01/2026 and the maximum flight time between planets is set at 2 years.
\item M2: the trajectory chosen for the JUICE mission to Jupiter \cite{Grasset2013} is of type Earth-Earth-Venus-Earth-Mars-Earth-Jupiter. The first Earth-to-Earth leg includes a deep-space $v_{\infty}$ leveraging manoeuvre, capability not available in our trajectory optimizer. Due to this limitation, the TPF-generated sequence (obtained from three $v_{\infty}$ contours at each planet) passed to the optimizer is
Earth-Venus-Earth-Mars-Earth-Jupiter, as shown in Fig.~\ref{fig:M3} (Earth 6, Earth 6 - Venus 6, Venus 6 - Earth 10, Earth 10 - Mars 10, Mars 10 - Earth 12, Earth 12 - Jupiter 6). From this series, the optimizer generates a 7-year trajectory departing on 19/03/2023 with $v_{\infty}$ of 4.9 km/s and performing powered flybys with Venus (28/10/2023, 84 m/s), Earth (08/08/2024, 148 m/s), Mars (14/02/2025, 1105 m/s) and Earth (16/11/2026, 394 m/s). The arrival $v_{\infty}$ at Jupiter is 5.6 km/s. The departure dates are varied from 01/01/2023 to 31/12/2023 and the flight time between planets is limited to 2 years.
\end{itemize}
The performance of TPF for these scenarios is summarized in Table~\ref{table:TPF_performance}. 


\begin{table}[ht]
  \center
  \caption{CPU time, number of encounter paths and number of planet sequences for the three validation cases (V1, V2, V3) and the two mission scenarios (M1, M2) discussed in the text.}
  \begin{tabular}{lrrr}
	\hline 
      &  CPU time  (s) & Number of       & Number of \\ 
			&             & encounter paths & planet sequences \\ \hline
  V1  & 0.940      & 3376            & 12               \\
  V2  & 110.000    & 124104          & 101              \\
  V3  &  0.060     & 154             & 7                \\
  M1  &  0.004     & 4               & 2                \\
  M2  &  1.300     & 854             & 41               \\ \hline
  \end{tabular}
  \label{table:TPF_performance}
\end{table}

\section{Conclusions}
\label{sec:conclusions}
We described the algorithm and underlying theory of an automated method (TPF) for extracting interplanetary paths from a Tisserand graph. The algorithm uses a depth-first tree search method, building an exhaustive collection of all the encounter sequences connecting the departure and arrival planets. The search is depth limited (i.e., there is a maximum number of flybys in the path) to keep the algorithm from creating long repetitive sequences of encounters (loops). The method identifies seamlessly those situations in which repeated encounters with the same planet are required to respect the minimum pericenter height constraint. The algorithm is completely automatic, relieving the user from the burden of visually identifying the contour intersections. The process is very fast in contemporary commodity hardware. Even the most complex scenarios can be analyzed in a matter of minutes (we presented a case with 124 000 distinct paths that completes in under 110 seconds). This efficiency gives the user additional freedom when choosing the number of excess velocity contours to inspect, because the impact on the duration of the analysis is very limited.
The automated inspection of the TG reduces the workload of the mission designer and eliminates the chance of human error, which is inherent to any tedious tasks. Furthermore, when TPF is coupled with a trajectory optimizer, the complete workflow of mission design and optimization can be streamlined.

\section*{Funding sources}
The work of E. Fantino and R. Flores has been supported by Khalifa University of Science and Technology's internal grants FSU-2018-07 and CIRA-2018-85. R. Flores also acknowledges financial support from the Spanish Ministry of Economy and Competitiveness, through the "Severo Ochoa Programme for Centres of Excellence in R\&D" (CEX2018-000797-S).

\bibliography{references}   

\begin{thebibliography}{32}
\newcommand{\enquote}[1]{``#1''}
\providecommand{\natexlab}[1]{#1}
\providecommand{\url}[1]{\texttt{#1}}
\providecommand{\urlprefix}{URL }
\expandafter\ifx\csname urlstyle\endcsname\relax
  \providecommand{\doi}[1]{doi:\discretionary{}{}{}#1}\else
  \providecommand{\doi}{doi:\discretionary{}{}{}\begingroup
  \urlstyle{rm}\Url}\fi

\bibitem[{Roy(2005)}]{Roy2005}
Roy, A.~E., \emph{{Orbital motion}}, 4\textsuperscript{th} ed., Institute of
  Physics Publishing, Bristol, UK, 2005.

\bibitem[{Strange and Longuski(2002)}]{Strange2002}
Strange, N.~J., and Longuski, J.~M., \enquote{{Graphical method for
  gravity-assist trajectory design},} \emph{Journal of Spacecraft and Rockets},
  Vol.~39, No.~1, 2002, pp. 9--16.
\newblock \doi{10.2514/2.3800}.

\bibitem[{Miller and Weeks(2002)}]{Miller2002}
Miller, J.~K., and Weeks, C.~J., \emph{{Application of Tisserand's criterion to
  the design of gravity assist trajectories}}, 2002.
\newblock \doi{10.2514/6.2002-4717}, paper AIAA 2002-4717.

\bibitem[{Heaton et~al.(2002)Heaton, Strange, Longuski, and
  Bonfiglio}]{Heaton2002}
Heaton, A.~F., Strange, N.~J., Longuski, J.~M., and Bonfiglio, E.~P.,
  \enquote{{Automated Design of the Europa Orbiter Tour},} \emph{Journal of
  Spacecraft and Rockets}, Vol.~39, No.~1, 2002, pp. 17--22.
\newblock \doi{10.2514/2.3801}.

\bibitem[{Diehl et~al.(1983)Diehl, Kaplan, and Penzo}]{Diehl1983}
Diehl, R.~E., Kaplan, D.~I., and Penzo, P.~A., \enquote{{Satellite Tour Design
  for the Galileo Mission},} \emph{21$^{st}$ Aerospace Sciences Meeting}, 1983.
\newblock \doi{10.2514/6.1983-101}.

\bibitem[{Heaton and Longuski(2003)}]{Heaton2003}
Heaton, A.~F., and Longuski, J.~M., \enquote{{The feasibility of a
  Galileo-style tour of the Uranian satellites},} \emph{Journal of Spacecraft
  and Rockets}, Vol.~40, No.~3, 2003.
\newblock \doi{10.2514/2.3981}.

\bibitem[{Okutsu and Longuski(2002)}]{Okutsu2002}
Okutsu, M., and Longuski, J.~M., \enquote{{Mars Free Returns via Gravity Assist
  from Venus},} \emph{Journal of Spacecraft and Rockets}, Vol.~39, No.~1, 2002,
  pp. 31--36.
\newblock \doi{10.2514/2.3778}.

\bibitem[{Khan et~al.(2004)Khan, Campagnola, and Croon}]{Khan2004}
Khan, M., Campagnola, S., and Croon, M., \emph{{End-to-End Mission Analysis for
  a Low-Cost, Two-Spacecraft Mission to Europa}}, 2004.
\newblock Paper AAS 041-132.

\bibitem[{Campagnola and Russell(2010{\natexlab{a}})}]{Campagnola2010a}
Campagnola, S., and Russell, R.~P., \enquote{{The Endgame problem Part 1:
  V-Infinity Leveraging technique and the Leveraging graph},} \emph{Journal of
  Guidance, Control, and Dynamics}, Vol.~33, No.~2, 2010{\natexlab{a}}, pp.
  463--475.
\newblock \doi{10.2514/1.44258}.

\bibitem[{Strange et~al.(2010)Strange, Campagnola, and Russell}]{Strange2009}
Strange, N.~J., Campagnola, S., and Russell, R.~P., \emph{{Leveraging flybys of
  low mass moons to enable an Enceladus orbiter}}, 2010.
\newblock Paper AAS 09-435.

\bibitem[{Campagnola et~al.(2010)Campagnola, Strange, and
  Russell}]{Campagnola2010C}
Campagnola, S., Strange, N.~J., and Russell, R.~P., \enquote{{A fast tour
  design method using non-tangent V-Infinity Leveraging Transfers},}
  \emph{Celestial Mechanics and Dynamical Astronomy}, Vol. 108, 2010, pp.
  165--186.
\newblock \doi{10.1007/s10569-010-9295-1}.

\bibitem[{Campagnola and Russell(2010{\natexlab{b}})}]{Campagnola2010B}
Campagnola, S., and Russell, R.~P., \enquote{{Endgame Problem Part 2: Multibody
  Technique and the Tisserand-Poincare Graph},} \emph{Journal of Guidance,
  Control, and Dynamics}, Vol.~33, No.~2, 2010{\natexlab{b}}, pp. 476--486.
\newblock \doi{10.2514/1.44290}.

\bibitem[{Campagnola et~al.(2012)Campagnola, Skerritt, and
  Russell}]{Campagnola2012}
Campagnola, S., Skerritt, P., and Russell, R.~P., \enquote{{Flybys in the
  planar, circular, restricted, three-body problem},} \emph{Celestial Mechanics
  and Dynamical Astronomy}, Vol. 113, 2012, pp. 343--368.
\newblock \doi{10.1007/s10569-012-9427-x}.

\bibitem[{Kloster et~al.(2011)Kloster, Petropoulos, and Longuski}]{Kloster2011}
Kloster, K.~W., Petropoulos, A.~E., and Longuski, J.~M., \enquote{{Europa
  orbiter tour design with Io gravity assists},} \emph{Acta Astronautica},
  Vol.~68, No. 7-8, 2011, pp. 931--946.
\newblock \doi{10.1016/j.actaastro.2010.08.041}.

\bibitem[{Lantoine et~al.(2011)Lantoine, Russell, and
  Campagnola}]{Lantoine2011}
Lantoine, G., Russell, R.~P., and Campagnola, S., \enquote{{Optimization of
  low-energy resonant hopping transfers between planetary moons},} \emph{Acta
  Astronautica}, Vol.~68, No. 7-8, 2011, pp. 1361--1378.
\newblock \doi{10.1016/j.actaastro.2010.09.021}.

\bibitem[{Hughes et~al.(2013)Hughes, Moore, and Longuski}]{Hughes2013}
Hughes, K.~M., Moore, J.~W., and Longuski, J.~M., \emph{{Preliminary Analysis
  of Ballistic Trajectories to Neptune via Gravity Assists from Venus, Earth,
  Mars, Jupiter, Saturn, and Uranus}}, 2013.
\newblock Paper AAS 13-805.

\bibitem[{Strange et~al.(2014)Strange, Landau, Chodas, and
  Longuski}]{Strange2014}
Strange, N., Landau, D., Chodas, P.~W., and Longuski, J.~M.,
  \emph{{Identification of Retrievable Asteroids with the Tisserand
  Criterion}}, 2014.
\newblock \doi{10.2514/6.2014-4458}, paper AIAA 2014-4458.

\bibitem[{Colasurdo et~al.(2014)Colasurdo, Zavoli, Longo, Casalino, and
  Simeoni}]{Colasurdo2014}
Colasurdo, G., Zavoli, A., Longo, A., Casalino, L., and Simeoni, F.,
  \enquote{{Tour of Jupiter Galilean moons: Winning solution of GTOC6},}
  \emph{Acta Astronautica}, Vol. 102, 2014, pp. 190--199.
\newblock \doi{10.1016/j.actaastro.2014.06.003}.

\bibitem[{Campagnola et~al.(2014)Campagnola, Buffington, and
  Petropoulos}]{Campagnola2014a}
Campagnola, S., Buffington, B.~B., and Petropoulos, A.~E., \enquote{{Jovian
  tour design for orbiter and lander missions to Europa},} \emph{Acta
  Astronautica}, Vol. 100, No.~1, 2014, pp. 68--81.
\newblock \doi{10.1016/j.actaastro.2014.02.005}.

\bibitem[{Maiwald(2016)}]{Maiwald2016}
Maiwald, V., \enquote{{About Combining Tisserand Graph Gravity-Assist
  Sequencing with Low-Thrust Trajectory Optimization},} \emph{6$^{th}$
  International Conference on Astrodynamics Tools and Techniques}, 2016.

\bibitem[{Y{\'{a}}rnoz et~al.(2016)Y{\'{a}}rnoz, Yam, Campagnola, and
  Kawakatsu}]{Yarnoz2016}
Y{\'{a}}rnoz, D.~G., Yam, H., Campagnola, S., and Kawakatsu, Y.,
  \enquote{{Extended Tisserand-Poincare Graph and Multiple Lunar Swingby Design
  with Sun Perturbation},} \emph{6$^{th}$ International Conference on
  Astrodynamics Tools and Techniques}, 2016.

\bibitem[{Jones et~al.(2017)Jones, Hern\'andez, and Jesick}]{Jones2017}
Jones, D.~R., Hern\'andez, S., and Jesick, M., \emph{{Low Excess Speed Triple
  Cyclers of Venus, Earth, and Mars}}, 2017.
\newblock Paper AAS 17-577.

\bibitem[{Kaplan(1976)}]{Kaplan1976}
Kaplan, M.~H., \emph{{Modern Spacecraft Dynamics and Control}}, John Wiley \&
  Sons, 1976.

\bibitem[{Murray and Dermott(2000)}]{Murray2000}
Murray, C.~D., and Dermott, S.~F., \emph{{Solar System Dynamics}}, Cambridge
  University Press, Cambridge, UK, 2000.
\newblock \doi{10.1017/CBO9781139174817}.

\bibitem[{Curtis(2014)}]{Curtis2014}
Curtis, H.~D., \emph{{Orbital Mechanics for Engineering Students}},
  Butterworth-Heinemann, Oxford, UK, 2014.

\bibitem[{Dahlquist and Bjorck(2003)}]{Dahlquist2003}
Dahlquist, G., and Bjorck, A., \emph{{Numerical Methods}}, Dover Publications,
  Inc., USA, 2003.

\bibitem[{Dowell and Jarratt(1971)}]{Dowell1971}
Dowell, M., and Jarratt, P., \enquote{{A modified regula falsi method for
  computing the root of an equation},} \emph{Bit}, Vol.~11, No.~2, 1971, pp.
  168--174.
\newblock \doi{10.1007/BF01934364}.

\bibitem[{Cormen et~al.(1990)Cormen, Leiserson, Rivest, and Stein}]{Cormen1990}
Cormen, T.~H., Leiserson, C.~E., Rivest, R., and Stein, C.~.,
  \emph{{Introduction to Algorithms}}, 2\textsuperscript{nd} ed., MIT Press,
  1990.
\newblock \doi{10.1017/CBO9781107415324.004}.

\bibitem[{Korf(1985)}]{Korf1985}
Korf, R.~E., \enquote{{Depth-first iterative-deepening: An optimal admissible
  tree search},} \emph{Artificial intelligence}, Vol.~27, 1985, pp. 97--109.

\bibitem[{{De La Torre} et~al.(2020){De La Torre}, Fantino, Flores,
  Garc{\'{i}}a, and Calvente}]{DelaTorre2020}
{De La Torre}, D., Fantino, E., Flores, R., Garc{\'{i}}a, C., and Calvente, O.,
  \enquote{{Tisserand PathFinder Algorithm},}
  \url{https://gitlab.upc.edu/juno/tpf}, 2020.

\bibitem[{{The JUICE Science Working Team}(2014)}]{EuropeanSpaceAgencyESA2014}
{The JUICE Science Working Team}, \enquote{{Jupiter Icy moons Explorer
  Exploring the emergence of habitable worlds around gas giants},} Tech. rep.,
  {European Space Agency}, 2014.
\newblock ESA/SRE(2014)1.

\bibitem[{Grasset et~al.(2013)Grasset, Dougherty, Coustenis, and {et
  al.}}]{Grasset2013}
Grasset, O., Dougherty, M., Coustenis, A., and {et al.}, \enquote{JUpiter ICy
  moons Explorer (JUICE): An ESA mission to orbit Ganymede and to characterise
  the Jupiter system,} \emph{Planetary and Space Science}, Vol.~78, 2013, pp. 1
  -- 21.
\newblock \doi{10.1016/j.pss.2012.12.002}.

\end{thebibliography}

\end{document}